\documentclass{emulateapj}

\usepackage{rotating}
\usepackage{subfigure}
\usepackage{captcont}
\usepackage{amsmath}
\usepackage{calc}

\newcommand\aastex{AAS\TeX}
\def\h2{H$_2$}
\def\-2e{$^{-2}$\ }
\def\-2{$^{-2}$}

\newcommand{\HI}{\ensuremath{\mbox{\rm \ion{H}{1}}}}
\newcommand{\HII}{\ensuremath{\mbox{\rm \ion{H}{2}}}}

\renewcommand{\t}[1]{\mathrm{#1}}

\newcommand{\htwo}{\ensuremath{\mbox{H$_2$}}}
\newcommand{\msun}{\ensuremath{M_\odot}}

\newcommand{\sunits}{\mbox{\msun ~pc$^{-2}$}}
\newcommand{\pc}{\ensuremath{\mbox{pc}}}
\newcommand{\kms}{\mbox{km~s$^{-1}$}}

\newcommand{\cm}{\mbox{cm$^{-2}$}}

\newcommand{\co}[1]{\mbox{$^{#1}$CO}}

\newcommand{\counits}{\mbox{K km s$^{-1}$}}

\newcommand{\vunits}{\mbox{km s$^{-1}$ pc$^{-1}$}}
\newcommand{\junits}{\mbox{pc km s$^{-1}$}}
\renewcommand{\d}[1]{\ensuremath{\mbox{d}{#1}}}
\newcommand{\arc}{\mbox{$^{\prime\prime}$}}

\begin{document}

  \slugcomment{Accepted for Publication in ApJ}

   \title{Angular Momentum in Giant Molecular Clouds. I. The Milky Way}

   \author{Nia Imara and Leo Blitz}
   \affil{Astronomy Department, University of California, Berkeley, CA 94720}
   \email{imaran@berkeley.edu}
  
  \begin{abstract}
     We present a detailed analysis comparing the velocity fields in molecular clouds \emph{and} the atomic gas that surrounds them in order to address the origin of the gradients. To that end, we present first-moment intensity-weighted velocity maps of the molecular clouds and surrounding atomic gas.
     The maps are made from high-resolution \co{13} observations and 21-cm observations from the Leiden/Argentine/Bonn Galactic \HI~Survey.
     We find that (i) the atomic gas associated with each molecular cloud has a substantial velocity gradient---ranging within 0.02 to 0.07 \vunits---whether or not the molecular cloud itself has a substantial linear gradient.  (ii) If the gradients in the molecular and atomic gas were due to rotation, this would imply that the molecular clouds have less specific angular momentum than the surrounding \HI~by a factor of 1 -- 6.  (iii) Most importantly, the velocity gradient position angles in the molecular and atomic gas are generally widely separated---by as much as $130^\circ$ in the case of the Rosette Molecular Cloud.  This result argues against the hypothesis that molecular clouds formed by simple top-down collapse from atomic gas.
   \end{abstract}


\section{Introduction}\label{sec:intro} 
Giant molecular clouds, both Galactic and extragalactic, are observed to have velocity gradients that many authors have interpreted as being caused by rotation (e.g., Kutner et al. 1977, Phillips 1999; Rosolowsky et al. 2003).  If we start with the premise that these clouds are rotating because they have inherited the angular momentum of the rotating galactic disk out of which they formed, conservation of angular momentum should provide clues that give us insight to the origin of their formation.  However, simple formation theories  that assume giant molecular clouds (GMCs) form by condensing out of the Galactic disk are at odds with some of the observations.  For instance, they do not adequately explain why the directions of GMC velocity gradients are not typically aligned with the direction of Galactic rotation, counter to the expectation of conservation of angular momentum. Furthermore, simple formation scenarios tend to overpredict the observed specific angular momentum o
f GMCs (e.g., Blitz 1990; Rosolowsky et al. 2003).  This is the so-called ``angular momentum problem.''  

Provided there is no transfer of angular momentum, the angular momentum of a GMC should be equal to that of the gas out of which it formed. But Blitz (1990), working under the assumption that the velocity gradients in molecular clouds are due to rotation, showed that the angular momentum due to Galactic differential rotation in the solar neighborhood interstellar medium is consistently greater than that contained within molecular clouds.  Even molecular clouds with the largest observed velocity gradients---such as the Rosette and Orion A molecular clouds---have less specific angular momentum compared to the ISM from which they presumably formed.  Blitz (1990) also pointed out that because the molecular clouds in his sample are rotating in a sense \emph{opposite} to that of Galactic rotation, they could not have conserved angular momentum from the initial states he calculated, unless the local Galactic rotation curve is falling, or unless the clouds always collapsed azimuthall
y.

The angular momentum problem also extends to extragalactic molecular clouds. Roso-lowsky et al. (2003) showed that simple GMC formation theories consistently overestimate the magnitude of the observed angular momentum of molecular clouds in the galaxy M33.   On average, they found that simple theory overpredicts the observed magnitudes of specific angular momentum by more than a factor of 5.  Furthermore, they found that 40\% of the GMCs in M33 are counterrotating with respect to the galactic plane. 

In this paper, we shed light on the angular momentum problem by doing a detailed analysis of the kinematics in GMCs \emph{and} the surrounding ISM.  Whereas previous studies estimated the initial angular momentum imparted to GMCs from the Galactic rotation curve (e.g., Blitz 1990; Blitz 1993), we compare the velocity fields of GMCs to those of the ISM with which they are associated directly from observation.  In light of the observation that GMCs have a spatial and kinematic correlation with high-surface density atomic gas (\S \ref{sec:analysis}), we pose the question: Does the rotation of the large-scale \HI~associated with GMCs mirror that of the GMCs themselves?  Our primary goal is to determine whether or not rotation is the cause of the velocity gradients in GMCs.  To that end, we create first-moment maps of the molecular clouds in our sample and of the atomic gas surrounding them are for comparison.  In the following section, we describe the \co{13} and 21-cm data used 
to conduct this study.  In \S 3, we create intensity-weighted first moment \co{13} and \HI~maps of five Galactic clouds: Perseus, Orion A, NGC 2264, Monoceros R2 (MonR2), and the Rosette.  The results from these measurements are given in \S 4, and a summary is provided in \S 5.

\section{Data}\label{sec:data}
To measure velocity gradients and other properties across the molecular clouds, we use high-resolution, high-sensitivity published \co{13} observations.   Because \co{13} emission is nearly always optically thin in Galactic GMCs, we have the advantage of getting a detailed view of the kinematic structure of the molecular clouds in our sample.  And because \co{13} has narrower line widths than the optically thick \co{12}, the former permits finer separation of velocity components than the latter.  Nevertheless, the large scale velocity gradient of a GMC measured using \co{13} is generally consistent with that measured using \co{12}, since the gradient is being measured across several parsecs and small scale variations in the velocity field tend to get averaged out.   

 The data for Perseus, Orion A, NGC 2264, and  the MonR2 molecular clouds were generously provided by J. Bally (see Bally et al. 1987).   Observations of these clouds were taken at the AT\&T Bell Laboratories 7 m telescope and have a beam size of 100\arc.  Perseus, Orion A, and NGC 2264 data were resampled onto $60\arc$ grids.  The MonR2 data were resampled onto a $30\arc$ grid.   The spectral resolution of 128 channels at 100 kHz corresponds to a velocity resolution of 0.27 \kms.  On the $T_A^\ast$ scale, the cubes have rms noise levels of 0.17 K (Perseus), 0.32 K (Orion A), 0.42 (NGC 2264) and 0.29 K (MonR2).  

J. Williams and M. Heyer graciously provided the FCRAO data of the Rosette Molecular Cloud (see Heyer, Williams, \& Brunt 2006).   The beam size is $47\arc$ and the data were interpolated onto a $20\arc$ grid.  The  spectral resolution is 59 kHz per channel, and the velocity resolution is 0.133 \kms.  The data have an rms noise of 0.21 K in $T_A^\ast$, similar to that of the Bell Labs data.

The \HI~data are obtained from the Leiden/Argentine/Bonn (LAB) Galactic \HI~Survey (Kalberla et al. 2005), which spans velocities from $-400$ \kms~to $+400$ \kms.  (The LSR velocities of the molecular clouds in our sample range from 6 to 14 \kms~with velocity dispersions of a few \kms.)  The survey has a half power beam width of $0.6^{\circ}$, velocity resolution of 1.3 \kms, and an rms noise level of 0.07 K.  The high sensitivity and resolution of the LAB data set enables a detailed study of the atomic gas from which the GMCs formed.

\section{Analysis}\label{sec:analysis}
Observations in the Milky Way indicate the molecular clouds are typically associated with high-density atomic gas with column densities around $N(\HI)\sim 2\times 10^{21}\cm$ (e.g., McKee \& Ostriker 2007).  The Rosette Molecular Cloud is a prototypical example: Williams et al. (1995) measured the mean column of the \HI~associated with the GMC to be $1.3\times 10^{21}\cm$.   Furthermore, in external galaxies, GMCs are often observed to be located on or near bright \HI~peaks.  In the Large Magellanic Cloud, Mizuno et al. (2001) observed that most GMCs are associated with \HI~having column densities greater than $10^{21}$ \cm.  Rosolowsky et al. (2003) found that every GMC they identified in M33 lies on an over-dense \HI~filament, though every over-dense \HI~region does not contain a GMC.  This seems to imply that high column density atomic gas is necessary but not sufficient for GMC formation.

It is our goal to do a detailed comparison of the velocity fields in the GMCs and local atomic gas.  To that end, we first describe how the physical properties---including column densities, masses, and velocity gradients---in both the atomic and molecular gas are determined.  We then provide our criteria for choosing spatial and kinematic regions of atomic gas associated with the molecular clouds.  Lastly, in this section, we discuss how we estimate the specific angular momentum.  These parameters are summarized in Tables 1 and 2.

\subsection{Cloud Properties}
 The \HI~column density, $N(\HI)$, is calculated along each line of sight by integrating the atomic hydrogen emission above a certain background value and over the selected velocity range, (see \S 3.1 and Table 1), 
\begin{equation}\label{eq:NHI}
   N(\t{\HI})=1.82\times 10^{18}~\int_{v_\t{min}}^{v_\t{max}} \frac{T_{\t{b,HI}}}{\counits}~\d v~\cm,
\end{equation}
where $T_{\t{b,HI}}$ is the brightness temperature of the \HI~observations and $\d v$ is the channel velocity width.  Assuming that the \HI~is optically thin, Equation \ref{eq:NHI} provides a lower limit to measured column density, which we convert into units of surface density in Figures \ref{fig:oriona0} -- \ref{fig:rosette0}.

The total \HI~mass is then determined by summing over all pixels in the map where emission is detected, that is, where the emission is at least three times the root-mean-square (rms) noise level:
\begin{equation}
  M_{\t{HI}}=\sum_{\t{pixels}} \mu \cdot m_{\t{H}}\cdot N(\t{\HI})_{\t{pixel}}\cdot(d^2\Delta\alpha\Delta\delta),
\end{equation}
where $\mu=1.36$ is the correction for helium, $d$ is the distance to a given molecular cloud, and $d^2\Delta\alpha\Delta\delta$ is the area of one pixel, which corresponds to one resolution element.
 
The \co{13} column density, $N(\co{13})$, is derived assuming that the \co{13}~emission is optically thin and in local thermodynamic equilibrium.  Following Frerking et al. (1982),
    \begin{eqnarray}\label{eq:Nco13}
      N(\co{13})=2.13\times 10^{14}~ [1-\t{e}^{-5.287/ T_{\t{ex}}}]^{-1}  \nonumber \\
        \times \int_{v_\t{min}}^{v_\t{max}}{\frac{ T_{\t{b,CO}}~\d v}{\counits}} ~\cm,
    \end{eqnarray}
where $\int T_{\t{b,CO}}~\d v$ is the integrated \co{13} intensity and $T_{\t{ex}}$ is the excitation temperature.  Normally, $T_{\t{ex}}$ is determined by measuring the \co{12} radiation temperature toward \co{13} peaks.  Since we lack \co{12} observations at the same resolution as the \co{13} data, we use a constant excitation temperature in the calculation of  $N(\co{13})$ for each of the five GMCs.  Based on the following arguments,  we adopt a value of 20 K for each of the GMCs.  If the actual excitation temperature in a given region is between 10 and 30 K, the derived \co{13} column density will be in error by less than a factor of 2.  

Castets et al. (1990) showed that the \co{13} emission in Orion A mainly arises from regions where $T_{\t{ex}} \approx 20-25$ K, and dense cloud cores have temperatures of $T_{\t{ex}} \approx 15-20$ K.  Nagahama et al. (1998) showed that, with the exception of two peaks at $l\sim 209^\circ$ and $l\sim 212.5^\circ$ associated with embedded young stellar groups,  $T_{\t{ex}}$ rises slowly and monotonically in Galactic longitude, with an average ranging from $13-20$ K.  In his study of molecular clouds, including Perseus, Orion A, and MonR2, Carpenter (2000) adopts a constant value of  $T_{\t{ex}}=10$ K, though the coefficient in his formula for $N(\co{13})$ yields slightly higher values than ours in Equation \ref{eq:Nco13}.  Thus, for Orion A, Perseus, NGC 2264, and Monoceros R2, we use $T_{\t{ex}}=20$ K in our calculation of $N(\co{13})$.

Williams et al. (1995) showed that $T_{\t{ex}}$ in the Rosette decreases slowly with increasing distance from the Rosette Nebula (centered at $l=206.2^\circ,~b=-2.1^\circ$) from $\sim 20$ to $5$ K.  In our analysis, we adopt a uniform value of $T_{\t{ex}}=20$ K for the Rosette.  Our estimate of the cloud's mass using this value is slightly lower than that estimated by Williams et al. (1995), who measured the mass over a larger surface area (see below).

Next, the \htwo~column density is evaluated assuming a ratio of $N(\htwo)/N(\co{13})=7\times 10^5$ (Frerking et al. 1982).  Pixels having values at least three times the rms noise level are counted as detected.  Finally, the molecular mass $M_{13}$ is calculated over the areas where emission is detected (that is, higher than the 3-$\sigma_\t{rms}$ level) using
    \begin{equation}\label{eq:mco}
  M_{13}=\sum_{\t{pixels}} \mu \cdot m_{{\t H}_2} \cdot N(\htwo)_{\t{pixel}}\cdot(d^2\Delta\alpha\Delta\delta),
    \end{equation}
where  $m_{{\t H}_2}$ is the mass of an \htwo~molecule. We note that the areas over which emission is detected and, subsequently, the masses we calculate will be smaller than cited in previous studies in which these quantities were measured using \co{12} emission.  This is because the stronger \co{12} line is observed over larger areas in GMCs than the \co{13} line.  In the Rosette, for instance, we measure a projected area of 1500 $\pc^2$ and a mass of $6.0 \times 10^4$ \msun, while Williams et al. (1995) measure 2200 $\pc^2$ and $7.7 \times 10^4$ \msun.  In Table 1, both the \co{13} masses and previously measured \co{12} masses are listed for the GMCs.

\subsection{Velocity Gradients}\label{subsec:gradients}
Velocity gradients are measured from first moment maps of the atomic and molecular gas.  First, the intensity-weighted average velocity along each line of sight is determined using
\begin{eqnarray}
 v_\t{lsr}=\frac{\sum_{i} v_i T_i}{\sum_{i} T_i},
\end{eqnarray}
where $v_i$ and $T_i$ are the velocity and temperature at location $i$.  Following Goodman et al. (1993), the uncertainty of a given measurement is
\begin{eqnarray}
\sigma_{\t{lsr}} = 1.2\left(\frac{T_\t{rms}}{T_\t{peak}}\right)(\d v ~\Delta v_{\t{FWHM}})^{1/2},
\end{eqnarray}
where $T_\t{rms}$ is the spectrum noise, $T_\t{peak}$ is the maximum temperature along the line of sight, and $\Delta v_{\t{FWHM}}$ is the FWHM linewidth of the spectrum along the line of sight.

A plane is then fitted to the first moment map of velocity centroids, as in Goodman et al. (1993), assuming a linear velocity gradient:
\begin{equation}
   v_{\t{lsr}} = v_0 + a (x-x_0) + b(y-y_0),\label{eq:vlsr}
\end{equation}
where $v_0$ is the mean cloud velocity, and $(x_0,y_0)$ is an arbitrary reference position, which we take to be the center of our maps, and the coefficients are
\begin{equation}
   a=\frac{\partial v}{\partial x}, ~~b=\frac{\partial v}{\partial y}.
\end{equation}
The gradient magnitude and direction, $\Omega$ and $\theta$, are derived from the coefficients to the fit,
   \begin{equation}
     \Omega\equiv |\nabla v_{\rm{lsr}}|=\frac{(a^2+b^2)^{1/2}}{d}, \label{eq:omega}
   \end{equation}
   \begin{equation}
     \theta = \tan^{-1}\frac{b}{a},  \label{eq:theta}
   \end{equation}
where $\theta$, measured in degrees East from North, points in the direction of increasing velocity.  Note that since we have no information regarding the inclination of a given cloud, $i$, to our line of sight, gradient measurements are underestimates of the actual values, $\Omega_\t{true}=\Omega/\sin i$. 

The uncertainties in these values are calculated by propagating the errors in the coefficients. To check whether planes are good fits to the velocity centroid maps, we make plots of the central velocity at a given location in the cloud versus the perpendicular offset from the cloud's rotation axis.  This is done by taking the average velocity along lines parallel to the rotation axis at various distances.  In most cases, as will be discussed in \S \ref{sec:results}, these plots show that planes are good fits to the velocity centroid maps.

\subsection{Selecting \HI~Regions}
We select \HI~regions in the position-velocity LAB data cube that are centered, spatially and kinematically, on the five GMCs in our sample.  We do not know \emph{a priori} the kinematic or spatial extent of the \HI~regions that are associated with each GMC nor the extent of the \HI~velocity gradients.  Molecular clouds have well-defined boundaries at which the molecules are dissociated by UV radiation and where there is a distinct transition from primarily molecular to primarily atomic gas (e.g. Savage et al. 1977; Blitz \& Thaddeus 1980).  The atomic gas associated with GMCs does not have such distinct boundaries, however,  making it difficult to distinguish \HI~that may be related to GMCs from background emission.  Thus, we begin by examining regions in position-velocity space that are far from the center of the GMC.  While we want to capture the full extent of any linear velocity gradient we may measure in a given region of atomic gas, we do not want to make our aperture 
so large that we end up including in our measurements too much atomic gas that is unrelated to the molecular clouds.  

We start by varying the size of the region (that is, the subcube extracted from the LAB data)  and examine how the velocity gradient magnitude and direction change.  We fix the velocity range (see below) and vary the spatial size of the region centered on the GMC from about 10 to several tens of parsecs, in increments of 10 pc.  We find that the gradient direction remains roughly constant until the radius of the region over which it is measured reaches $40\pm 10$ pc, independent of the size of the GMC.  Beyond this, the measurements start to fluctuate, as the gradient in the vicinity of the molecular cloud becomes washed out by unrelated features.  

\begin{figure*}[htbp]
\centering
\includegraphics[scale=.65]{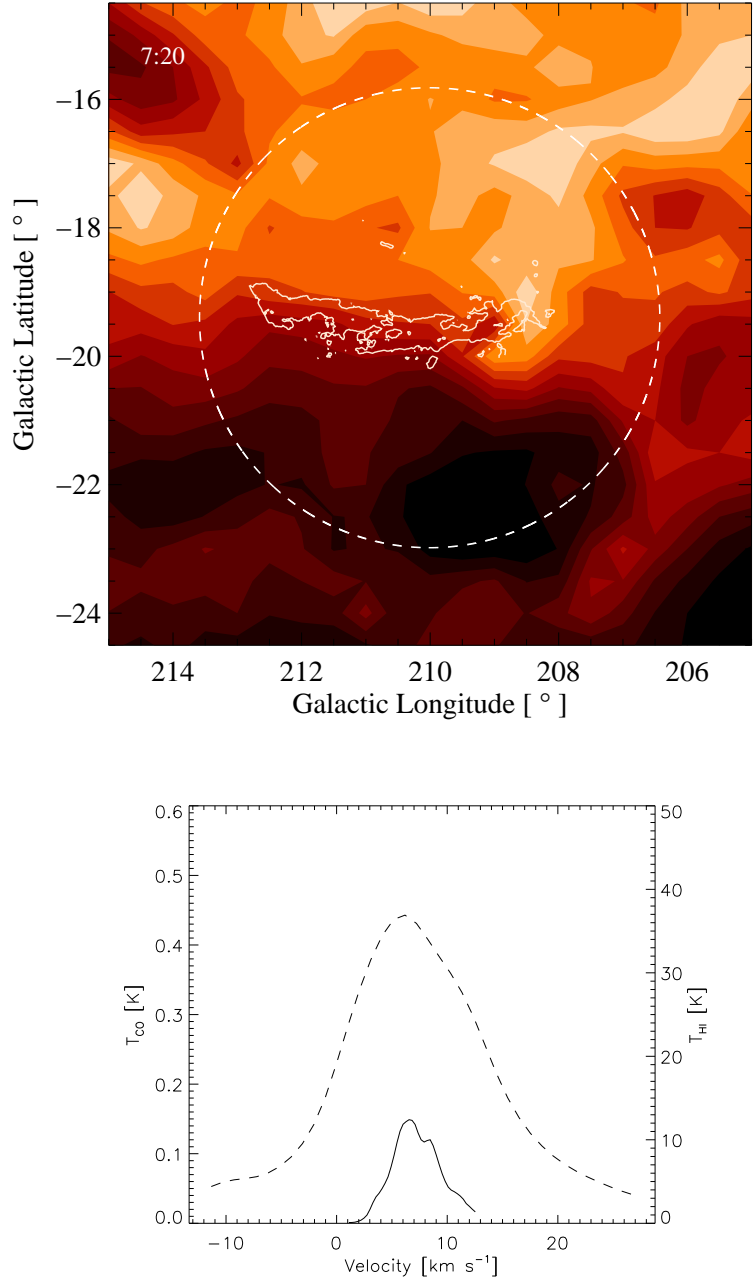}
\vspace{-1.5em}
\caption{Orion A. The bottom figure plots the average spectra of the \co{13} emission (solid line) in the GMC and of the \HI~emission (dashed line) in the region of atomic gas.  The top figure shows a surface density map of \HI~overlaid with an outline of the molecular cloud.  The range of \HI~surface density, in units of \sunits, is marked in the top left-hand corner, and the contour spacing is 1 \sunits.  Atomic gas within the dashed line is used for the subsequent analysis described in the text. \label{fig:oriona0}}
\end{figure*}

\begin{figure*}[htbp]
\centering
\includegraphics[scale=.7]{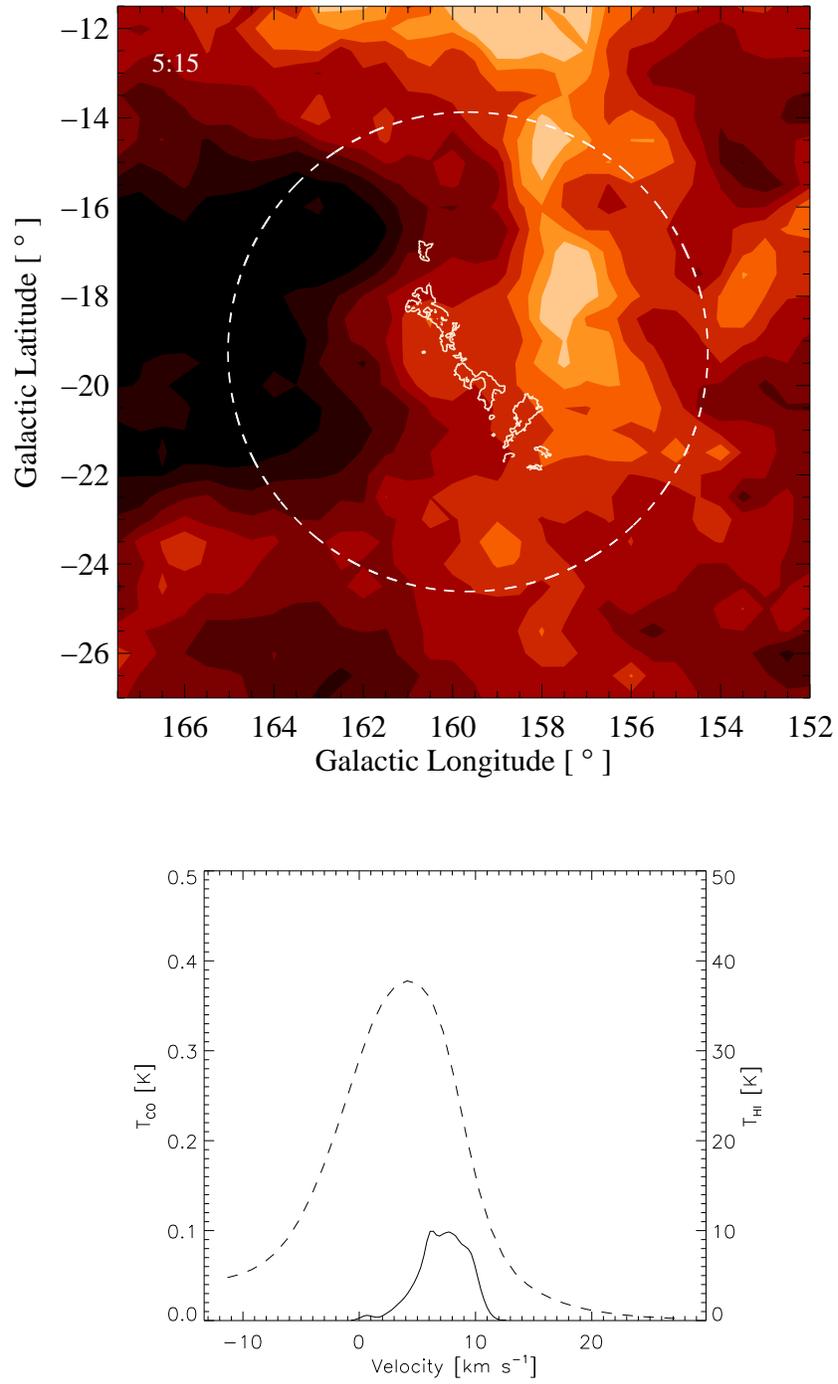}
\caption{Perseus.  Same as Figure \ref{fig:oriona0}.\label{fig:perseus0}}
\end{figure*}

\begin{figure*}[htbp]
\centering
\includegraphics[scale=.7]{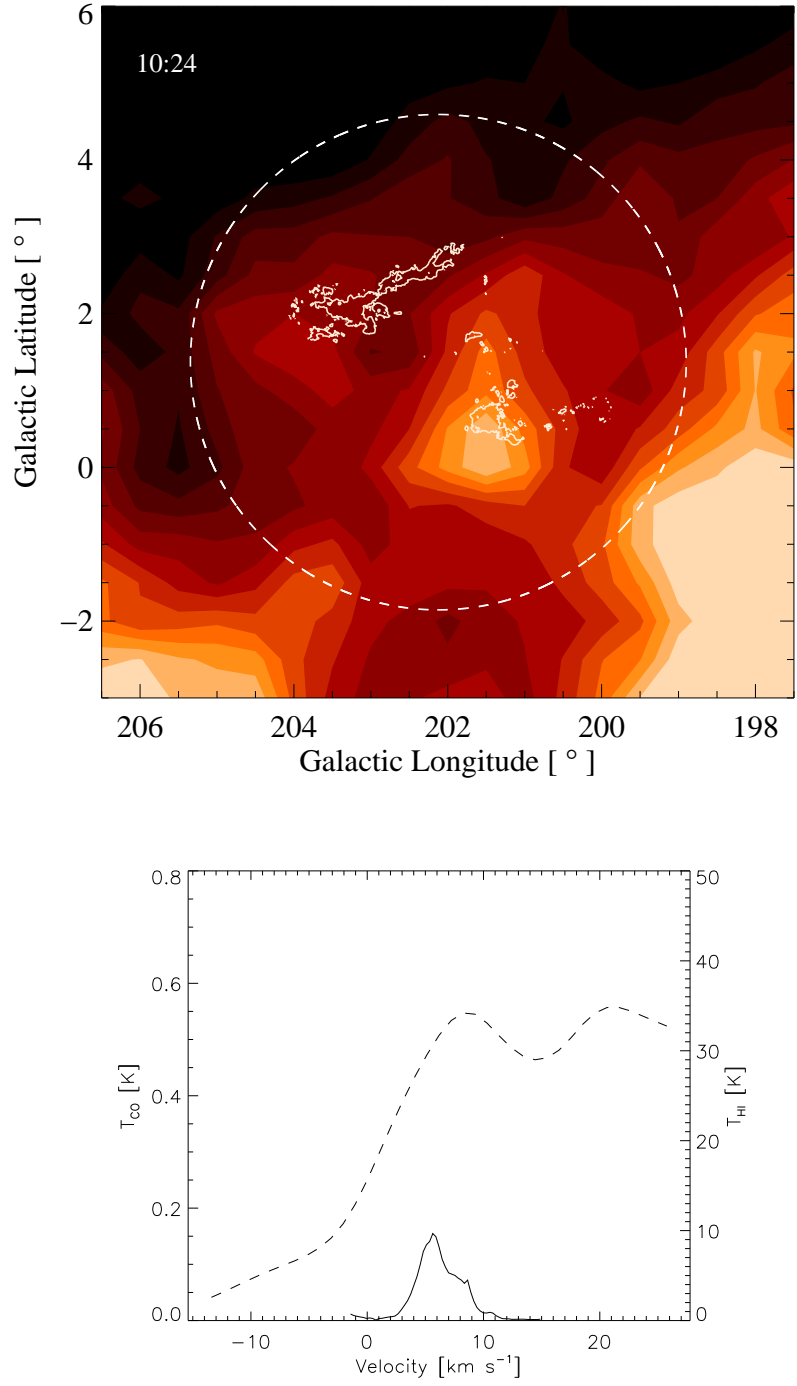}
\caption{NGC 2264. Same as Figure \ref{fig:oriona0}.  \label{fig:ngc22640}}
\end{figure*}

\begin{figure*}[htbp]
\centering
\includegraphics[scale=.7]{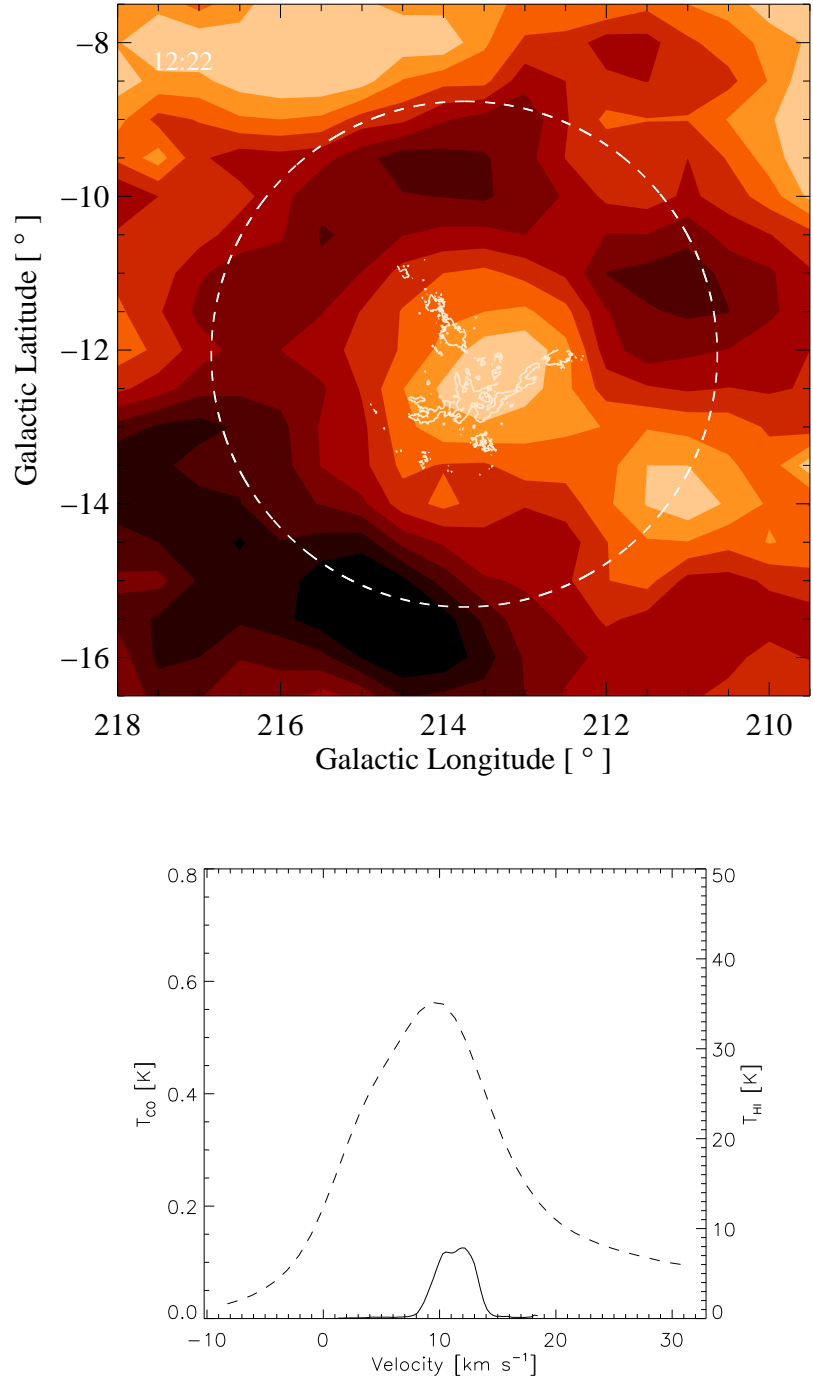}
\caption{MonR2. Same as Figure \ref{fig:oriona0}.  \label{fig:monr20}}
\end{figure*}

\begin{figure*}[htbp]
\centering
\includegraphics[scale=.7]{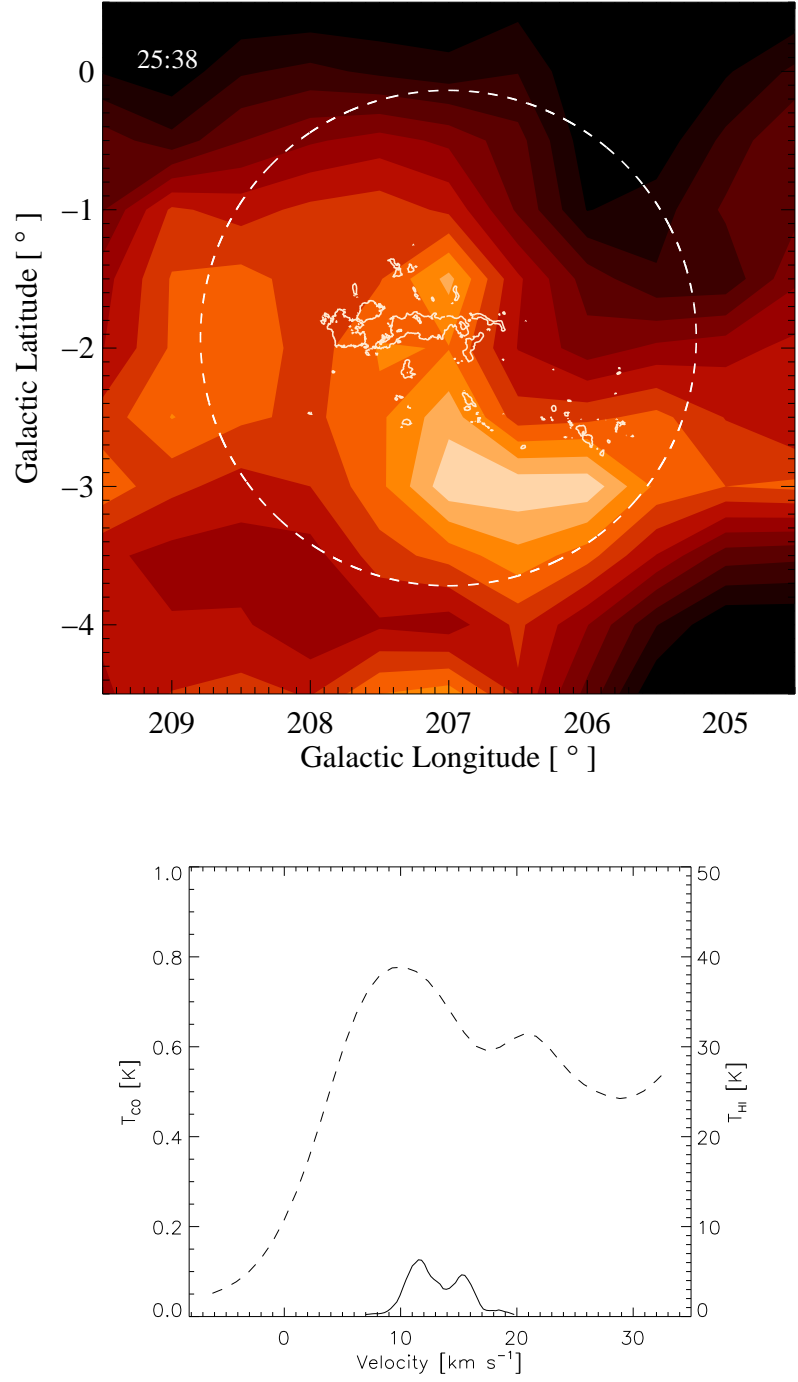}
\caption{The Rosette. Same as Figure \ref{fig:oriona0}. \label{fig:rosette0}}
\end{figure*}

Thus for each cloud, we end by selecting a spatial boundary of atomic gas which extends roughly 30 -- 50 pc from the center of the CO emission.  Because the peaks of the \HI~regions are included in the maps we generate, varying the size of the region within $40\pm 10$ pc does not change the gradient direction by more than a few degrees.  The 1-$\sigma$ uncertainty level of $\theta$ ranges from 3 to 6 degrees for the clouds in our sample.  Our criterion is supported by previous studies such as that of Andersson et al. (1991) who found that the spatial extent of high-intensity \HI~halos, measured from the edge of the molecular cloud, ranges 5 -- 10 pc (see also \S \ref{sec:discussion}).  Figures \ref{fig:oriona0} through \ref{fig:rosette0} show the \HI~surface density maps derived from the zeroth-moment intensity maps, with the outline of the molecular clouds overlaid.  Also overplotted in each figure is a dashed circle indicating the \HI~region selected for the analysis.  Keep
 in mind, we are showing the \co{13} emission of the molecular clouds, and so the maps  in Figures \ref{fig:oriona0} -- \ref{fig:rosette0} do not show the full extent of the molecular emission in the GMCs.

To choose relevant velocities of the atomic gas associated each GMC, we begin by examining the \HI~emission in the velocity range $\pm 20~\kms$ centered about the mean LSR velocity of the \co{13} emission.  Again, this is because we want to be sure that our measurements include as much as possible of the associated atomic gas.  In the direction of the Rosette, for instance, the \HI~has been observed to extend several \kms~beyond the CO emission (Williams et al. 1995).   Studies of both Milky Way molecular clouds (e.g., Wannier et al. 1983, Williams et al. 1995) and extragalactic clouds (e.g. Engargiola et al. 2003) have shown that the \HI~emission line profile tends to peak in the direction of GMCs.  In effect, we are using the \HI~velocity as a proxy for distance in order to associate the atomic gas with the GMCs.  As Figures \ref{fig:oriona0} -- \ref{fig:rosette0} show, although the \HI~emission line is broader than the CO line, the velocity difference between the peaks in 
the respective lines never exceeds $\sigma_\t{HI}$, where $\sigma_\t{HI}$ is the velocity dispersion of the \HI~profile.   

The bottom panel in Figures \ref{fig:oriona0} through \ref{fig:rosette0}  show the average \co{13} spectrum through each GMC with the \HI~spectrum in the same direction (within the dashed circle) overplotted.  Because the \co{13}--\HI~peaks are nearly coincident in each case, this indicates that most of the \HI~in the direction of a given cloud is associated with that cloud within the selected velocity range.  In the cases of Perseus, Orion A, and MonR2, the \HI~emission drops abruptly beyond $\pm 15~\kms$ of the \co{13}~emission.  NGC 2264 and the Rosette have more complicated \HI~spectra, each showing double peaks that may be indicative expansion or of a blended, possibly unrelated component.  The latter explanation would not be surprising since, of the five clouds in the sample,  NGC 2246 and the Rosette are located closest to the Galactic plane where line-of-sight blending is more of a problem.

Based on Figures \ref{fig:oriona0} -- \ref{fig:rosette0}, we determine the boundaries of the \HI~emission we will use for the subsequent analysis.  For Perseus, Orion A, and MonR2, the \HI~line profiles are approximated as Gaussians and we assume that  \HI~emission having velocities within $\pm 2\sigma_\t{HI}$ is associated with a given molecular cloud.  Since both NGC 2264 and the Rosette each have a second peak in their \HI~temperature profiles ($T_\t{b,HI}$) at higher velocities and because we want to be careful to exclude as much unrelated emission as possible, we set a slightly more stringent criterion on the velocities we select.  For each cloud, a maximum velocity is identified at the local minimum in $T_{\t{b,HI}}$ where the Gaussians overlap.  In the NGC 2264 spectrum, for instance, $T_{\t{b,HI}}$ drops to 29 K at 15 \kms~and then peaks again at around 20 \kms~(see Figure \ref{fig:ngc22640}).  Thus, we eliminate all emission having velocities above 15 \kms, the locat
ion of the local minimum.    All of these selections are listed in the third to last column of Table 1. 


\subsection{Specific Angular Momentum}\label{subsec:j}
Once the magnitudes of the velocity gradients in the molecular clouds and the surrounding \HI~are measured, we may calculate and compare their specific angular momenta, under the assumption that the linear gradients are due to solid body rotation.  The specific angular momentum, $j$, is simply the total angular momentum of a body divided by its mass,
\begin{equation}
j=\beta \Omega R^2, \label{eq:j}
\end{equation}
 where $R$ is the radius of the region, and the constant $\beta$ takes into account the moment of inertia of a rotating body.  For roughly spherical GMCs having constant surface mass density distributions, $\beta=2/5$.  Unless otherwise stated, we take the size of a given molecular cloud to be its effective radius, as defined by its projected area: $R_\t{eff}=\sqrt{A/\pi}$.

We would like to estimate the expected specific angular momentum initially imparted to a GMC by the ISM from which it forms.  This depends on the process of GMC formation, and we assume here that GMCs form via a ``top-down'' formation mechanism.  For instance, it has been suggested that molecular cloud formation occurs when an instability triggers collapse or condensation from the Galactic disk (e.g., Mouschovias et al. 1974; Blitz \& Shu 1980; Elmegreen 1982; Kim et al. 1998).  Blitz \& Shu (1980) show that ``bottom-up'' formation of GMCs via the random agglomeration of pre-existing low-mass clouds is unlikely because of the long timescales for this process.  Gravitational instabilities and magneto-gravitational instabilities, however, tend to proceed more quickly.

If the \HI~surrounding the GMCs is reflective of the ISM out of which the GMCs initially formed, the quantity $(1/2)\Omega R^2$ provides an estimate of the initial specific angular momentum of the GMCs, assuming they are rotating.  Henceforth, we will often refer to this as the \emph{expected} specific angular momentum.  Note that this disregards the possible effects of magnetic fields; that is, we are assuming that the magnetic field strength of the forming cloud is 0.  The initial angular velocity imparted to a given GMC is the local value of $\Omega$, which we calculate from the first-moment maps of the \HI~using Equation \ref{eq:omega}.  In principle, the size of the region from which a forming cloud gathers material, the ``accumulation radius'' $R_A$,  could have a range of values because it depends on details of the formation mechanism, as well as on assumptions regarding the initial surface density and geometry of the gas from which a GMC formed.  Nevertheless, we can 
set an effective lower limit on $R_A$ by calculating the size of the region from which a GMC must contract to obtain its present mass, $M_{\t{GMC}}$. 

Following Blitz (1993), let us assume that the initial geometry of the collapsing region is a cylinder with a diameter equal to its height.  The size of the cylinder is determined by requiring that the mass contained within it is equal to the present mass of the GMC:
\begin{equation}
2 \pi \rho_\t{HI} R_A^3 = M_\t{GMC},\label{eq:accum}
\end{equation}
where $\rho_\t{HI}$ is the  mean volume density of the atomic gas from which the GMC formed.  Blitz (1993) estimated $R_A$ using the mean value of $\Sigma_\t{HI}$ in the Galactic plane near the Sun, 5 \sunits~(Henderson et al. 1982).  This corresponds to a mass density of $\rho_{\t{HI}}= 0.0125 ~\msun~\pc^{-3}$ (or a number density of 0.5 cm$^{-3}$), using an effective scale height of atomic gas of 200 pc in the solar vicinity (Falgarone \& Lequeux 1973).  However, using larger values of $\Sigma_\t{HI}$ (leading to lower estimates of $R_A$) might be more appropriate, given the observation that GMCs tend to form in regions of \HI~with densities higher than global galactic values (e.g., Engargiola et al. 2003; Imara et al. 2010). We estimate the accumulation radii using $\Sigma_\t{HI}=10$ \sunits, a value more in keeping observations of the atomic gas associated with GMCs.  For instance, around the Rosette, Williams et al. (1995) measure an \HI~column density of $1.3\times 10^{
21}$ \cm, which corresponds to a $\Sigma_\t{HI}\approx 10$ \sunits, twice the mean Galactic value.  From the Sancisi (1974) study of atomic gas near Perseus, the inferred surface density is 11 \sunits.  Our estimates of $R_A$ are listed in Table 2, as well as the effective radii of the GMCs.


Finally, we note that  the predicted specific angular momentum in the atomic gas has a marked dependence on the GMC mass.  Substituting $R_A\sim (M_\t{GMC}/\rho_\t{HI})^{1/3}$ from Equation 2.12 for the radius in Equation 2.11 yields  $j_\t{HI}\sim M_\t{GMC}^{2/3}$. The masses we calculate using the \co{13} observations underestimate the total molecular mass of the GMCs.  For this reason, we  use the larger GMC masses, as previously measured using \co{12} observations, to predict the initial specific angular momentum imparted to forming molecular clouds.  Typically, $R_A\sim 3R_\t{GMC}$.


\section{Results}\label{sec:results}

\begin{figure*}[htbp]
\centering

\includegraphics[scale=.65]{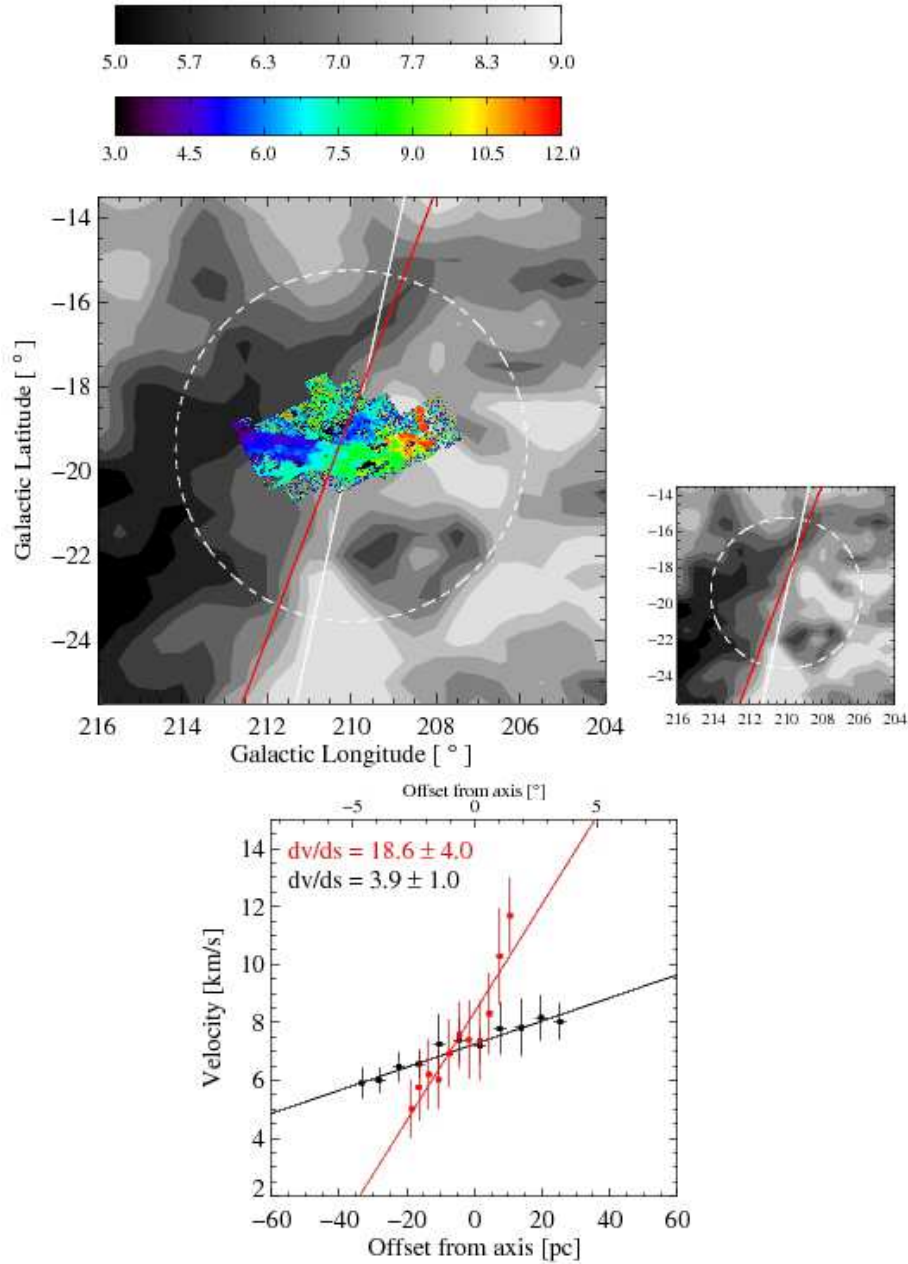}
\caption{Orion A.   The top figure shows the intensity-weighted first moment maps of \HI~(grayscale) overlaid with \co{13} (color).  Both maps have contour spacings of $0.5~\kms$, and the velocity ranges of the maps are indicated by the color bars.  The first moment map to the right shows the \HI~without the \co{13} overlaid.  The bottom figure plots the central velocity versus perpendicular offset from the rotation axis for pixels in the above map located within the dashed line.  The \co{13} data (in red) are binned every 3 pc and the \HI~data, every 6 pc.  The error bars indicate the dispersion within the bins.  \label{fig:oriona}}
\end{figure*}

\begin{figure*}[htbp]
\centering
\includegraphics[scale=.65]{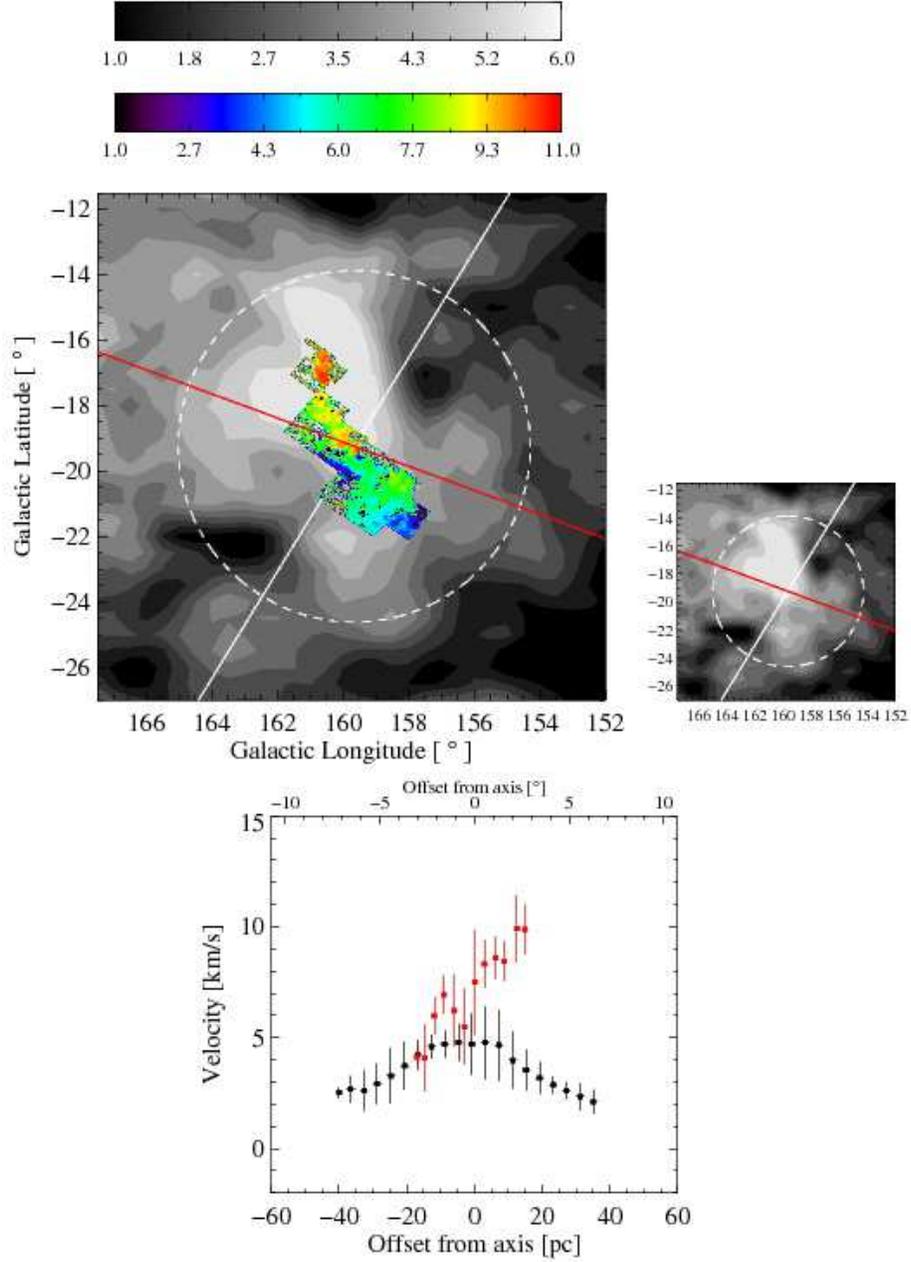}
\caption{Perseus.  Same as Figure \ref{fig:oriona}, except the \HI~data in the position-velocity plot are binned every 4 pc.  The non-linearity of the \HI~position-velocity plot indicates that there is not a linear velocity gradient over the entire region within the dashed circle.  In the following figure, we see that by changing the reference position, there is a significant linear gradient in the atomic gas. \label{fig:perseus}}
\end{figure*}

\begin{figure*}[htbp]
\centering
\includegraphics[scale=.65]{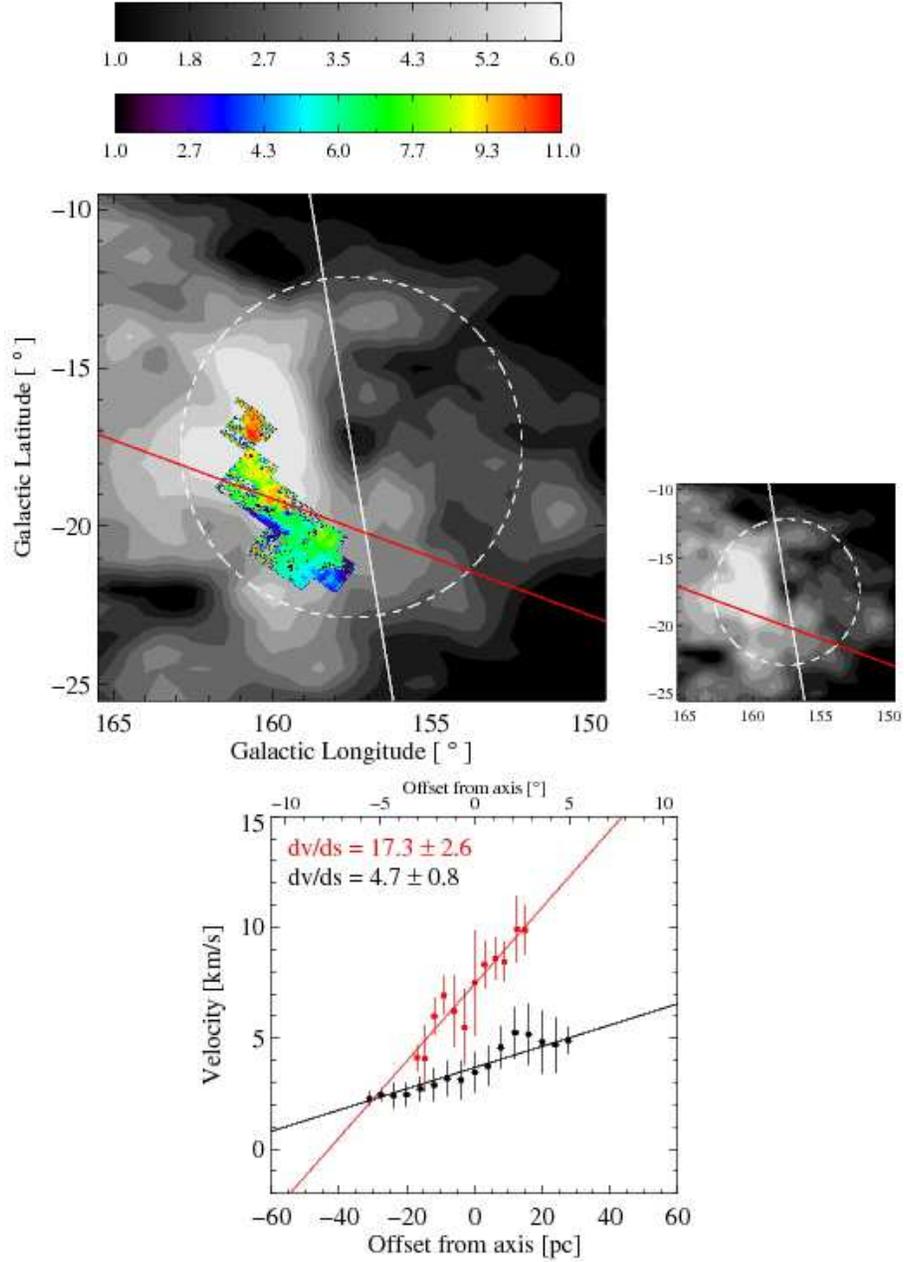}
\caption{Perseus.  Same as Figure \ref{fig:oriona}, accept the \HI data in the position-velocity plot are binned every 4 pc.  The surface density map in Figure \ref{fig:perseus0} shows that a high-density \HI peak is located to the West of the molecular cloud.  By changing the reference position to $l_0,b_0=157.5,-18$, near the center of the peak, we find that there is a significant linear gradient across the field centered on the peak.   \label{fig:perseus2}}
\end{figure*}

\begin{figure*}[htbp]
\centering
\includegraphics[scale=.65]{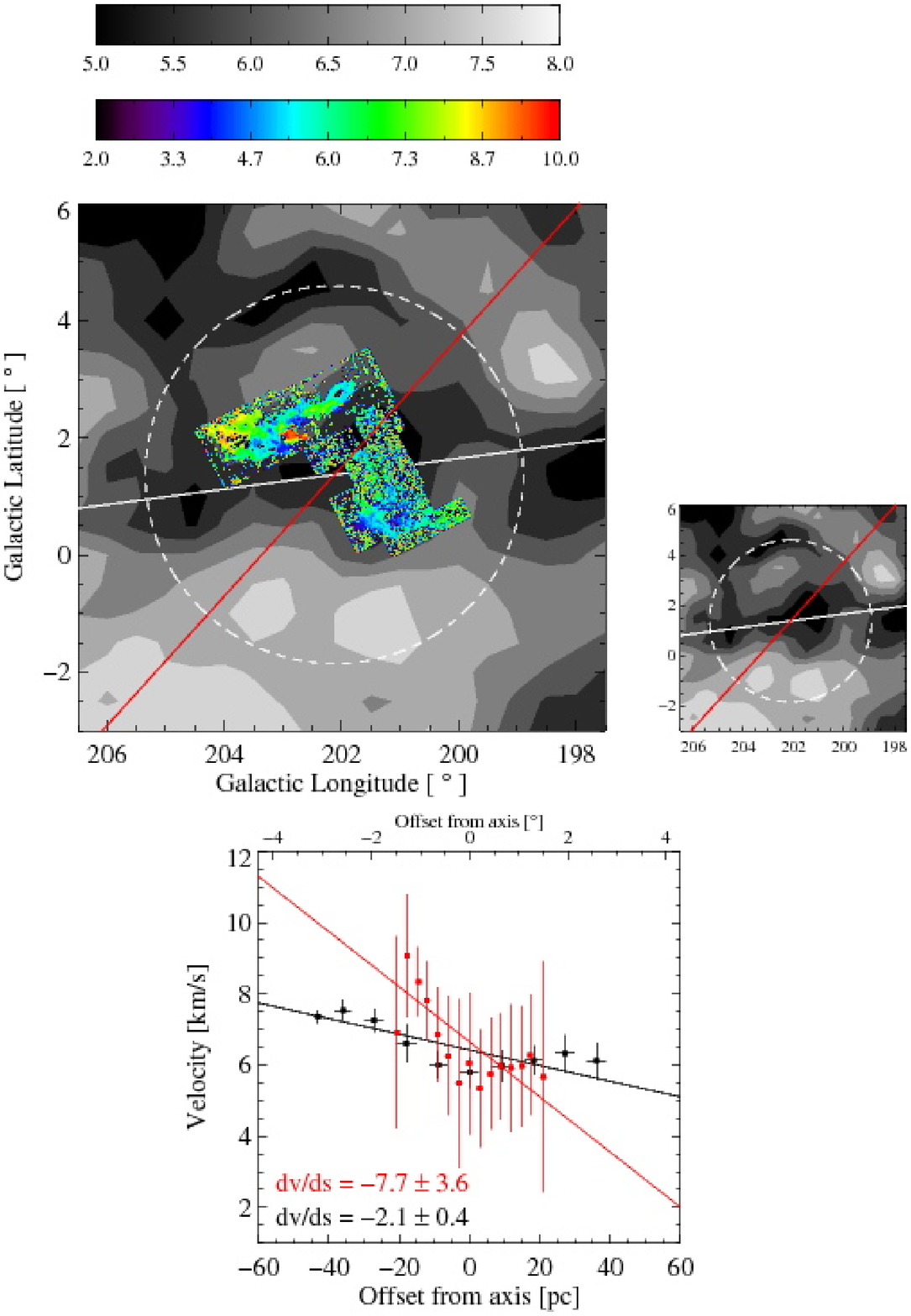}\\
\caption{NGC 2264. Same as Figure \ref{fig:oriona}, except the \HI~data in the position-velocity plot are binned every 9 pc.  Although there is no significant linear gradient across the face of the GMC, we plot the ``rotation axis'' (red line) measured using Equation \ref{eq:theta} as if there was. \label{fig:ngc2264}}
\end{figure*}

Figures \ref{fig:oriona} through \ref{fig:rosette} show the \HI~velocity maps (grayscale) overlaid with velocity maps of \co{13} (color) for each GMC.  Overplotted on these maps are axes of rotation: the lines perpendicular to the gradient directions, $\theta_\t{GMC}$ and $\theta_\t{HI}$, as calculated from Equation \ref{eq:theta}.  These lines are the position angles of the rotation axes, $\psi$, (where $\psi=\theta +90^\circ$), of the molecular and atomic gas \emph{if} the gradients are in fact due to rotation.  

In order to check whether planes are good fits to the first-moment velocity maps, we make position-velocity cuts parallel to the maximum gradient directions and plot the results (bottom panels of Figures \ref{fig:oriona} -- \ref{fig:rosette}).  We are essentially plotting the central velocity at a given location in the cloud versus displacement along the gradient on a pixel by pixel basis.  Figures \ref{fig:oriona}, \ref{fig:monr2}, and \ref{fig:rosette} show that planes are good fits to the first-moment maps of \HI~surrounding Orion A, MonR2, and the Rosette because there is a clear linear trend in the gradient in the atomic hydrogen.  The velocity fields of the \HI~associated with Perseus and NGC 2264 appear to have more complex structure.    In the case of the NGC 2264 and MonR2 molecular clouds (Figures \ref{fig:ngc2264} and \ref{fig:monr2}), however, the position-velocity plots do not have monotonically increasing or decreasing slopes, indicating a more complex velocity 
structure in the \emph{molecular} gas (red points).  In these two cases, we nevertheless overplot the lines perpendicular to the gradient direction calculated from Equation \ref{eq:theta}.

\begin{figure*}[htbp]
\centering
\includegraphics[scale=.4]{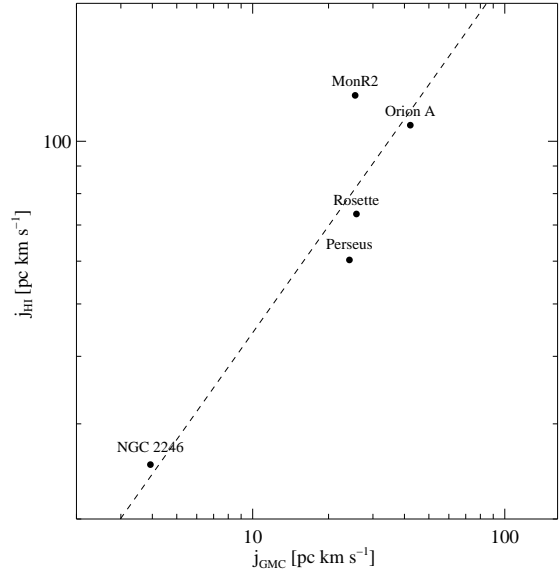}\\
\caption{Specific angular momentum of the atomic hydrogen, $j_\t{HI}$, versus GMC specific angular momentum, $j_\t{GMC}$. Overplotted is a least-squares fit to the data: $j_\t{HI}\propto j_\t{GMC}^{0.66\pm 0.20}$. \label{fig:jplot}}
\end{figure*}

\begin{figure*}[htbp]
\centering
\includegraphics[scale=.4]{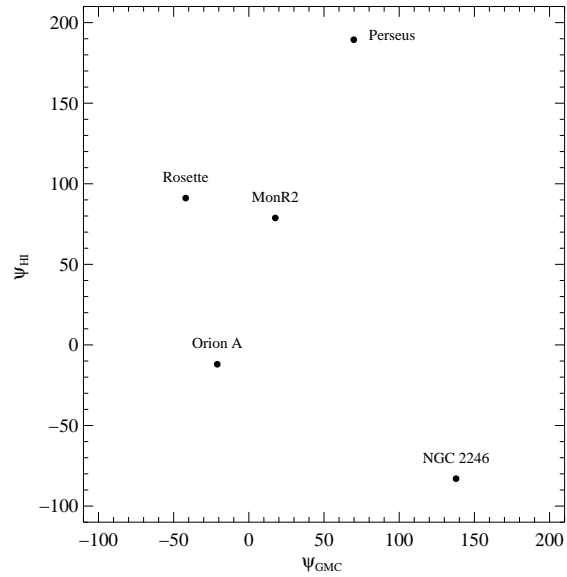}\\
\caption{The gradient position angles of the atomic hydrogen, $\psi_\t{HI}$, versus position angles of the GMCs, $\psi_\t{GMC}$ appear to be uncorrelated.  \label{fig:theta}}
\end{figure*}

The specific angular momenta in the atomic and molecular gas, listed in Table 3, are compared in Figure \ref{fig:jplot}.  There appears to be a reasonable correlation---$j_\t{HI}$ and $j_\t{GMC}$ increasing together as $j_\t{HI}\propto j_\t{GMC}^{0.66\pm 0.20}$---although small number statistics prevent us from making a firm conclusion.  In each case, the initially expected specific angular momentum, $j_\t{HI}$, is always greater than $j_\t{GMC}$.  These measurements alone are consistent with a picture whereby GMCs form via some top-down mechanism, such as a gravitational instability, and somehow shed angular momentum in the process.  But since angular momentum is a vector quantity, this scenario is difficult to reconcile with the  observation that the gradient position angles in the molecular and atomic gas differ and appear to be uncorrelated (Figure \ref{fig:theta}).

The key finding of our analysis is that the regions of atomic gas associated with molecular clouds have linear velocity gradients, yet the directions of these gradients are---with one exception---unaligned with the direction of the gradients in the associated GMCs.  Under the hypothesis that the gradients are caused by solid body rotation, this would imply that GMCs are \emph{not} corotating with the surrounding ISM.  The second key result is that the magnitudes of the velocity gradients in the GMCs are larger than the gradient magnitudes in the atomic gas. Thirdly, if the gradients in the molecular and atomic gas are from rotation, the angular momenta in the molecular clouds is less than predicted from calculations of the angular momenta in the associated atomic gas.  

Below, we describe the results in detail for each GMC.  The gradient directions, magnitudes, and specific angular momenta of both the molecular and atomic gas are given in Table 3.


\begin{center}  \textbf{\sc{Orion A}} \end{center}
Figure \ref{fig:oriona} shows that the Orion A molecular cloud has a large-scale gradient whose velocity decreases from about 12 \kms~to 3 \kms~with increasing Galactic longitude.  Of the GMCs in this study, observations of the kinematics and morphology of the Orion A molecular cloud seem to make the best case  for a top-down picture of GMC formation. The direction of the gradient and the long axis of Orion A are parallel to each other and to the Galactic plane.  If the gradient were due to rotation, this would imply that Orion A is rotating in a sense nearly opposite to that of the Galactic disk.  This result is in agreement with previous studies (e.g., Kutner et al. 1977, Blitz 1993).  The velocity gradient of the \HI~in the immediate surroundings of Orion A points in nearly the same direction as that in the molecular cloud.  Figure \ref{fig:oriona} shows that the gradient in the atomic hydrogen, integrated over the velocity range from about $-7$ to 22 \kms, differs from th
e gradient in the molecular cloud by only $9^\circ$.  

The origin of the velocity gradient in Orion A has been debated; it has previously been explained by cloud rotation (Kutner et al. 1977), expansion driven by the Orion OB association (Bally et al. 1987), and expansion driven by stellar winds from newborn stars (Heyer et al. 1992).  If due to rotation, the 0.22 \vunits~gradient in the molecular cloud implies a specific angular momentum of 42.2 \junits. However, considering that Orion A is filamentary and much more closely resembles a cylinder rotating about its minor axis than a sphere, this estimate of $j$ is likely to be low limit, since the configuration of the former has a higher moment of inertia than the latter.  For a cylinder rotating about its minor axis, $j_\t{cyl}=1/12\Omega(3R^2+L^2)$, where $R$ is the radius of the cylinder and $L$ is the length.  The cylindrical radius and length of Orion A are approximately $1.5^\circ$ and $9^\circ$, corresponding to 11 pc and 65 pc.  Consequently, $j_\t{cyl}=84$ \junits, much c
loser to, but still a bit less than the estimated angular momentum of the associated atomic gas (107 \junits).    

These two results---that the gradients of the molecular and local atomic gas are in near alignment, and the magnitudes of the specific angular momenta are within range of each other---suggest that Orion A may be a case in which the molecular cloud and surrounding \HI~are corotating and in which conservation of angular momentum is demonstrated.  As we will see below, it appears to be an exceptional case.

\begin{center} \textbf{\sc{Perseus}} \end{center}
Figure \ref{fig:perseus} shows that a strong linear gradient exists in the Perseus molecular cloud.  The velocities in the GMC range from about 1 to 11 \kms~with increasing Galactic latitude.  The gradient direction is tilted about $20^\circ$ to the North-West.  Of the molecular clouds in the sample, Perseus has the largest velocity gradient at 0.23 \vunits.   If the gradient is caused by rotation, from Equation \ref{eq:j}, the specific angular momentum in Perseus is 24.2 \junits, as indicated in Table 3.

Examination of the black points in the position velocity-plot shown in Figure \ref{fig:perseus} indicates that the velocity gradient of the \HI~region centered on Perseus is not linear.  As with each cloud in this study, we take the center of the \HI~field to be located at the position centroid of the GMC in Galactic coordinates, ($l_0,b_0$).  This affects the appearance of the position-velocity plot, which shows the velocity of points in the field as a function of distance from the gradient judged from $l_0,b_0$.  The \HI~surface density map in Figure \ref{fig:perseus0} shows a high-surface density ($\sim 16$ \msun) \HI~filament extending vertically near a Galactic longitude of $157.5^\circ$.  Whereas the other GMCs in this sample have some portion of their molecular material laying directly on top of an \HI~peak, this is not strictly the case with Perseus, which is tilted toward the East from the bright \HI~filament.  Thus, we generate another first-moment map of the \HI~ce
ntered on the filament at $l_0,b_0=157.5,-18$, as well as another position-velocity plot centered here. Figure \ref{fig:perseus2} shows that a linear fit is a much more suitable fit to the gradient in the \HI~when we shift the reference point.  When measured this way, the magnitude of the gradient measured using Equation \ref{eq:omega} shifts from 0.038 to 0.067 \vunits, the value we cite in Table 3.  The gradient direction is nearly $100^\circ$, very close to the sense of Galactic rotation, in which velocities increase from West to East.  But the directions of the gradients in the molecular cloud and the atomic gas surrounding the GMC differ by nearly $120^\circ$.   

Of the clouds in this analysis, Perseus has the largest velocity gradients in the molecular gas.  The magnitude of the gradient in the molecular gas (0.23 \vunits) is 3.4 times that in the \HI~($0.067$ \vunits).  The  estimated specific angular momentum in the atomic gas is $j_\t{HI}=60$ \junits, 2.5 times $j$ measured in the molecular cloud.  Although this is consistent with a scenario in which the Perseus molecular cloud formed by collapsing out of the atomic gas, its angular momentum being redistributed somehow in the process, based on the observation that there is such a large difference between the gradient directions in the molecular and atomic gas, it is difficult to see how the GMC could have formed in such a simple way.

\begin{center} \textbf{\sc{NGC 2264}} \end{center}
The first-moment maps and position-velocity plot in Figure \ref{fig:ngc2264} show that the NGC 2264 molecular cloud and surrounding \HI~both have complex kinematic features. Though there is no significant linear gradient across the entire GMC, there is a weak gradient in the atomic gas.  The gradient in the atomic gas is stronger to the South of the rotation axis, where the velocity decreases from about 8 to 5 \kms~as the Galactic latitude increases from $\sim -1.8^\circ$ to $+1.5^\circ$.  As seen in Figure \ref{fig:ngc22640}, the NGC 2264 molecular cloud is composed of two main structures.  The structure in the North has a larger range of velocities, with $v=2$ -- 10 \kms, than does the structure in the South, which averages around 6 \kms, close to the mean velocity of the molecular cloud as a whole.  This is also the average velocity of the \HI~peak located at near $l=201.5^\circ$, $b=0.5^\circ$ (Figure \ref{fig:ngc22640}), which is associated with the southern segment of t
he GMC.

The magnitude of the gradient in the atomic gas near NGC 2264 is 0.019 \vunits, which is less than $\Omega$($=0.025$ \vunits) in the solar vicinity assuming a flat rotation curve.  The magnitude of the gradient fitted across the entire GMC, using Equation \ref{eq:omega}, is 0.046 \vunits.  But the northern segment of the cloud in isolation has a gradient closer to 0.08 \vunits. 

It seems unlikely that rotation is the origin of the velocity field in the NGC 2264 molecular cloud.  Table 3 records the velocity magnitude and direction calculated by fitting a plane to the field, even though the gradient is not linear.  Using these values, the specific angular momentum of the GMC, $4$ \junits, is smaller than the expected value by at least a factor of 6.  The gradient direction in the GMC, pointing nearly $50^\circ$ East from North, differs from $\theta_\t{HI}$ by $140^\circ$.  An alternative explanation of the velocity field in NGC 2264 may be, at least in part, the internal stellar activity.  It is noteworthy that the \HII~region associated with NGC 2264 (the Cone Nebula) is located in the north, near the part of the GMC that has a larger velocity dispersion than the southern part.  The nebula, at $l=202.95$, $b=2.20$ (Kharchenko et al. 2005) is located between two high velocity regions of the GMC that are receding at speeds near 10 \kms.  This morpholog
y is suggestive of an expanding ring seen from an edge-on perspective.  Taken all together, it appears that these features could be causally connected: an \HII~region causes the high-speed expansion of the surrounding molecular gas, which subsequently sweeps up atomic gas into a high-density ridge.  It is perhaps significant that we see a similar pattern in the Rosette.

Yet another explanation of the kinematics of NGC 2264 is put forth by Fur\'{e}sz et al. (2006), based on their finding that the stars and \co{13} emission in NGC 2264 are well-correlated in position-velocity space.  They suggest that the velocity field of the GMC is explained by the models of Burkert \& Hartmann (2004), in which molecular clouds form from supersonic collisions of gas.  This argument seems to be corroborated by the overall pattern of the cloud's velocity structure: higher velocities at the outskirts of the cloud and lower velocities proceeding toward the center of the cloud.  This pattern could be explained by gravitationally driven infall at the interface of the colliding flows.  In this scenario, star formation occurs preferentially in the condensations that  develop during the collapse of the molecular cloud, thus explaining the position-velocity correlation observed between stars and high-density molecular gas.  

The colliding flows hypothesis may also explain the velocity structure of the atomic gas.  Figure \ref{fig:ngc2264} shows that the \HI~has slightly higher velocities at positions in the field far from the gradient axis and lower velocities close to the axis---again, suggestive of infall accelerated by gravity.  If other molecular clouds form by this mechanism, however, it is unclear why we do not see this pattern in the atomic gas surrounding the rest of the GMCs in our sample.  Possibly, the  NGC 2264 system still retains signs of its early formation history due to the relatively young age of the GMC ($\sim 1$ -- 3 Myr; Flaccomio et al. 2000; Ram\'{i}rez et al. 2004).  It is difficult to make a definitive conclusion since we are observing \HI~so close to the Galactic plane where the problem of line-of-sight blending is exacerbated.

\begin{center} \textbf{\sc{MonR2}} \end{center}

Figures \ref{fig:monr20} and \ref{fig:monr2} show that the MonR2 molecular cloud sits right on top of a high-density cloud of \HI~that has a strong linear velocity gradient.  The gradient, which extends for over 80 pc, has a magnitude of 0.035 \kms~and increases from about 7 \kms~to 12 \kms~from South to North-West.  If the gradient is due to rotation, $j=47$ -- 180 \junits, depending on the size of the region considered.  From conservation of angular momentum, one might expect MonR2 to have a gradient somewhere between $\Omega =0.38$ -- 0.84 \vunits, pointed about $11^\circ$ West of North like the gradient in the atomic gas.  This is not what we see in Figure \ref{fig:monr2}, however.

The velocity field of the MonR2 molecular cloud has a much more complex structure than does the atomic gas.  The outer northern, western, and southern edges of the cloud reach velocities up to 12 -- 13 \kms.  The inner portion of the cloud is moving at slower speeds around 9 -- 11 \kms, which is also the velocity range at which the \HI~peak located at $l=213.5^\circ$, $b= -12.5^\circ$ dominates.  Considering the apparently random nature of the velocity field, it is unlikely that the GMC is undergoing large-scale, coherent rotation. The position-velocity plot for the molecular cloud in Figure \ref{fig:monr2} (red points) has a zero slope, as does NGC 2264, indicating that it has no significant linear gradient.  It is difficult to say whether or not the present velocity field originated during the cloud's formation or during some later stage in its evolution. If the GMC originally had a more organized velocity field and did \emph{not} inherent its present field during formation
, perhaps a series of interactions with external forces or internal events---such as turbulence and star formation activity---have washed out any signature of a systematic velocity gradient which was previously present in the molecular cloud.

\begin{center} \textbf{\sc{The Rosette}} \end{center}

The separation between the gradient directions in the Rosette molecular cloud and the surrounding atomic gas is approximately $130^\circ$.  Figure \ref{fig:rosette} shows that in the velocities of the molecular cloud tend to increase from roughly 7 -- 14 \kms~in the North-East to as high as $\sim 17$ \kms~in the South and South-West.  The gradient in the \HI~is directed perpendicular to the Galactic plane.  We note that, of the clouds in our sample, the \HI~maps near the Rosette probably suffer the greatest degree of line-of-sight blending.  The cloud is very close to the Galactic plane, and at the distance of the Rosette, 1600 pc, we have the lowest spatial resolution ($\sim 17$ pc) in the atomic gas.    Therefore the surface density and velocity maps displayed in Figures \ref{fig:rosette0} and \ref{fig:rosette} almost certainly fail to capture  many of the local, small-scale variations in the structure of the atomic gas.

We measure a gradient in the molecular cloud of 0.09 \vunits, consistent with the 0.08 \vunits~measured by Williams et al. (1995).  If the gradient is due to rotation, the specific angular momentum of the cloud is approximately 26 \junits.  Based on the estimate of the minimum value of the accumulation radius, the specific angular momentum in the surrounding ISM is nearly 3 times larger.  
 
Figure \ref{fig:rosette0} shows the outline of the molecular cloud overlaid on a surface density map of the atomic gas, which is integrated over the range $v=4$ -- 27 \kms.  A ``shell'' of \HI~containing two high-density peaks is associated with the molecular cloud.  The brightest peak of the shell sits at the southern edge of the molecular cloud near a latitude of $b=-3^\circ$, and another peak in the surface density occurs near $b=-1.5^\circ$.  The southern portion of the \HI~shell appears to mimic the ring-like structure in the molecular cloud sitting just to the north of it.  When the location of the structures in both the molecular and atomic gas are compared with the first-moment map in Figure \ref{fig:rosette}, we see that the southern portion of the \HI~shell is moving near the same velocity, $\sim 13$ - 14 \kms, as the south-east segment of the ring in the GMC.  The western half of the ring in the GMC is moving at higher velocities.  Also note how the \HI~peak locate
d near $b=-1.5^\circ$ is moving slightly faster ($\sim 14$ \kms) than the molecular gas in that region ($\sim 13$ \kms).  All of this suggests that both the \HI~shell and the ring-like structure in the GMC are expanding.  Kuchar \& Bania (1993) demonstrated that the \HII~region NGC 2244 (the Rosette Nebula) could have given rise to the expansion in the atomic gas.  And Williams et al. (1995) found that certain properties of clumps within the GMC vary with distance from the nebula, centered at $l=206.25^\circ$, $b=-2.11$ (Celnik 1983).  

Taken all together, the evidence leads us to suggest that the gradient in the Rosette molecular cloud is \emph{not} caused by rotation, but by the high-luminosity \HII~region, NGC 2244.  High-energy winds of stars in NGC 2244 may have excavated a hole in the GMC, causing \HI~to be swept up into a high-density ridge by the expanding molecular gas (e.g., Kuchar et al. 1991, Kuchar \& Bania 1993).  In this picture, because the nebula has had less of an impact on distant regions of the cloud, these distant regions are moving at lower velocities compared to molecular gas near the \HII~region.

\section{Implications for GMC Formation}\label{sec:discussion}

Our key findings are that the angular momentum in the GMCs is less than that in the surrounding atomic gas, and  the velocity gradient position angles in the molecular and atomic gas are widely divergent---with Orion A being the one exception.  This leads us to suggest that rotation may \emph{not} be the best explanation of the velocity fields observed in the GMCs.  

Traditionally, at least three possible solutions have been invoked to resolve the angular momentum problem:

1. One or more of the assumptions in the theory are inappropriate. For instance, if the average surface density of the precursor atomic gas is underestimated, this will lead to an overestimate in the accumulation radius and, consequently, an overestimate of the angular momentum initially imparted to a GMC.  In our analysis, we used 10 \sunits~for the mean surface density of the precursor gas, twice the average value in the Solar vicinity.  Assuming that the initial gas had an even higher density, say 20 \sunits, this would not change the main result, namely that $j_\t{HI}$ is consistently greater than $j_\t{GMC}$.  This is because the accumulation radius depends weakly on the initial surface density of the gas ($R_A\propto \Sigma_\t{HI}^{-1/3}$).  On the other hand, if collapsing molecular clouds do not gather material far from the Galactic plane, using larger values of $R_A$ might be appropriate.  Yet this would only exacerbate the discrepancy between the predicted and obser
ved angular momenta.  It might also be argued that using the effective radius of a GMC leads to underestimates in $j_\t{CO}$ if the GMC is filamentary.  In the previous section we took this into consideration with Orion A and recalculated  $j_\t{CO}$ assuming a cylindrical morphology.  This raised $j_\t{CO}$, which, nevertheless, remained less than $j_\t{HI}$.   Moreover, varying the size of the GMC or the region from which it gathers material does not solve the problem of the gradient directions in the molecular and atomic gas being unaligned.

2. An alternative explanation of the angular momentum problem is that though the assumptions regarding the initial conditions may be valid,  there may be some kind of external braking force which rapidly reduces the angular momentum of a GMC during its initial condensation.  Magnetic braking is often evoked as an angular momentum shedding mechanism (e.g., Mouschovias 1977; Fleck \& Clark 1981; Mestel \& Paris 1984; Rosolowsky et al. 2003).  Zeeman splitting of the OH 18-cm line and the 21-cm line of neutral hydrogen have been used to measure the magnetic field strengths of GMCs.  Since GMCs are magnetized and MHD effects are expected to play a significant role in their evolution, magnetic braking---in which magnetic field lines anchoring a GMC to the ambient ISM provide the tension necessary to slow down rotation---is a possible solution to the angular momentum problem.  Heiles \& Troland (2005) measured the mean magnetic field strength in the cold neutral medium of the Milky
 Way to be $B_0=6\pm 1.8~\mu G$.  The braking time is set by the time it takes for Alfv\'{e}n waves to travel across a region of gas having a moment of inertia comparable to the GMC (Mestel \& Paris 1984).  In order for magnetic braking to be efficient at slowing down cloud rotation, it would have to occur on timescales no greater than the timescale for cloud collapse.  For gas having an initial density of 1 cm$^{-3}$ (=0.025 \msun~$\pc^{-3}$), the Alfv\'{e}n speed is $B_0(4\pi\rho)^{-1/2}\approx  9.5$ \kms.  In a region having an accumulation radius of $70$ pc, this corresponds to a braking time of roughly $7.4$ Myr.  By comparison, the timescale for self-gravitational cloud formation in a region having the same density, assuming that this process occurs on a timescale close to the dynamical free-fall time (Mouschovias \& Paleologou 1979; Mestel \& Paris 1984; Elmegreen 2007),  is $(3\pi/4G\rho)^{1/2}\approx 44$ Myr. 

However, the effectiveness of magnetic braking largely depends on how the mass of the  forming cloud compares to its magnetic critical mass.  Mestel \& Paris (1984) argue that braking will efficiently slow down rotation only if the mass, $M$, of the forming cloud is much less than its magnetic critical mass, $M_C$.  Using Crutcher's (1999) magnetic field strength measurements of molecular clouds, McKee \& Ostriker (2007) infer that GMCs have approximately $M>2M_C$, that is, they  are magnetically supercritical. Elmegreen (2007) argues that the ISM out of which GMCs form has a magnetic field that is ``near-critical,'' i.e., $M\sim M_C$.   Furthermore, magnetic braking slows rotation most efficiently when the cloud's angular momentum vector is perpendicular to the magnetic field (Mouschovias \& Paleologou 1979).  Due to the complexity of making precise measurements of the magnetic field in the ISM, it is not established whether or not this is the case.  Clearly, more observatio
ns are needed in order to determine the importance of magnetic fields in GMC evolution.


%

3. Finally, perhaps GMCs are not rotating after all, and the ``problem'' is unfounded.  As some authors have pointed out, the interpretation that velocity gradients indicate rotation is not unique.  Expansion and shear, for instance, also produce velocity gradients.  The numerical simulations of Hennebelle et al. (2008) suggest that the converging flows of atomic gas could produce GMCs.  If shearing occurs at the interface of colliding flows, might this result in an excess of shear, that is, values of $\Omega$ that are higher than the shear arising from the local Galactic rotation curve?  Assuming a flat rotation curve, the local value of $\Omega$ in the solar neighborhood is $0.025$ \vunits.  This is slightly less than, but basically comparable to the gradient magnitudes we measure in Perseus, Orion A, and MonR2 (Table 3).  As discussed previously, we are underestimating $\Omega$ in every case, since we have no information regarding the inclination of clouds.  Also, since bl
ending of emission along the line-of-sight is most extreme in the cases of NGC 2264 and the Rosette, which are situated close to the Galactic plane, it is likely that $\Omega_\t{HI}$ is underestimated to an even greater degree in these cases.  Nevertheless, we cannot conclude whether or not shear is the cause of the velocity gradients based on these observations.  It is an issue we would like to further investigate with a larger sample of clouds. 

\begin{figure*}[htbp]
\centering
\includegraphics[scale=.4]{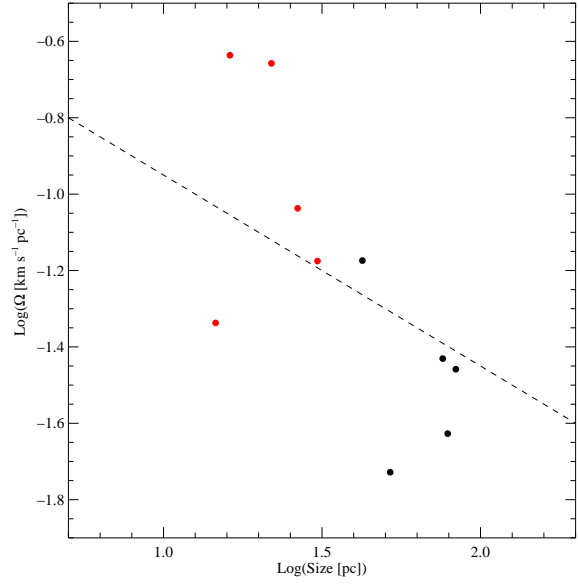}\\
\caption{The gradient magnitude versus size for \HI~and GMCs (red points).  Overplotted is the proportionality $\Omega\propto R^{-0.5}$ that Burkert \& Bodenheimer (2000) found for turbulent cores. \label{fig:r_vgrad}}
\end{figure*}

Burhert \& Bodenheimer (2000) show that turbulence may also cause linear velocity gradients.  They demonstrated that the gradient magnitude of turbulent cores scales with size as $\Omega\propto R^{-0.5}$.  Figure \ref{fig:r_vgrad} shows $\Omega$ as a function of $R$ for both the molecular clouds and \HI. Neither the GMCs nor the \HI~appear to follow the $\Omega\propto R^{-0.5}$ relationship, though we do not have enough data points to make a definite conslusion.

Studying the properties of GMC kinematics in other galaxies will also help to develop our picture of GMC formation.  In their analysis of 45 GMCs in M33, Rosolowsky et al. (2003) did the first systematic, extragalactic study of GMC angular momentum properties.  They showed that simple GMC formation theories consistently overestimate the magnitude of the observed angular momentum.  They measured the velocity gradients from high-resolution \co{12}($J=0\rightarrow 1$) data, finding that the gradients of M33 clouds are similar in magnitude to Galactic clouds.  They then tested several formation models by calculating the accumulation radii of the cataloged clouds which are predicted by the various models, including the Toomre and Parker instabilities.   On average, the theories, which do not include the effects of magnetic fields on rotation, overpredict observed magnitudes of velocity gradients by more than a factor of 5. And 40\% of the GMCs in M33 are counterrotating with respe
ct to the sense of galactic rotation. 

We extend this study by performing the analysis established here on a much larger sample of GMCs in M33 cataloged by Rosolowsky et al. (2003).  Since M33 has a relatively low inclination of $\sim 51^\circ$ (Corbelli \& Salucci 2000), such a study has the advantage of bypassing the problem of source confusion along lines of sight that arises when doing comparable surveys of molecular clouds in the Milky Way.

\section{Summary and Conclusions}\label{sec:summary}

We have presented a detailed comparison between the kinematics of five Galactic molecular clouds and the atomic gas that surrounds them.  We developed a method for selecting regions of \HI~that are associated with the GMCs and found that  each GMC was nearby high-density \HI~which peaked above the mean Galactic value.   First-moment maps were created using the \co{13} emission of the GMCs and the \HI~21-cm emission, and then a plane was fit to each map of velocity centroids.  We determined the magnitudes and directions of the velocity gradients from the coefficients to the fits.  From these observations and measurements, we arrived at the following conclusions:

1. Orion A, Perseus, and the Rosette each have a significant linear velocity gradient across the face of the molecular clouds, while NGC 2264 and MonR2 have complex, non-linear velocity fields.  The Perseus molecular cloud has the strongest linear gradient in the sample, with a magnitude of $0.23$ \vunits.

2. The atomic gas associated with Orion A, MonR2 and the Rosette has significant linear velocity gradients, regardless of whether the molecular cloud has one.   The \HI~gradients range from 0.019 to 0.067 \vunits, or 0.76 to 2.7 times the shear in the solar vicinity as measured by the Galactic rotation curve.  

3. If the gradients in the molecular and atomic gas were due to rotation, the specific angular momentum in the GMCs is less than that predicted by the formation scenario in which a GMC preserves angular momentum while undergoing top-down collapse from the surrounding ISM by a factor of 1 to 6.  The discrepancy can be narrowed if different assumptions are made regarding the initial density and geometry of the gas.  But the direction of the trend---that the observed angular momentum is less than the predicted---remains the same.

\begin{figure*}[htbp]
\centering
\includegraphics[scale=.65]{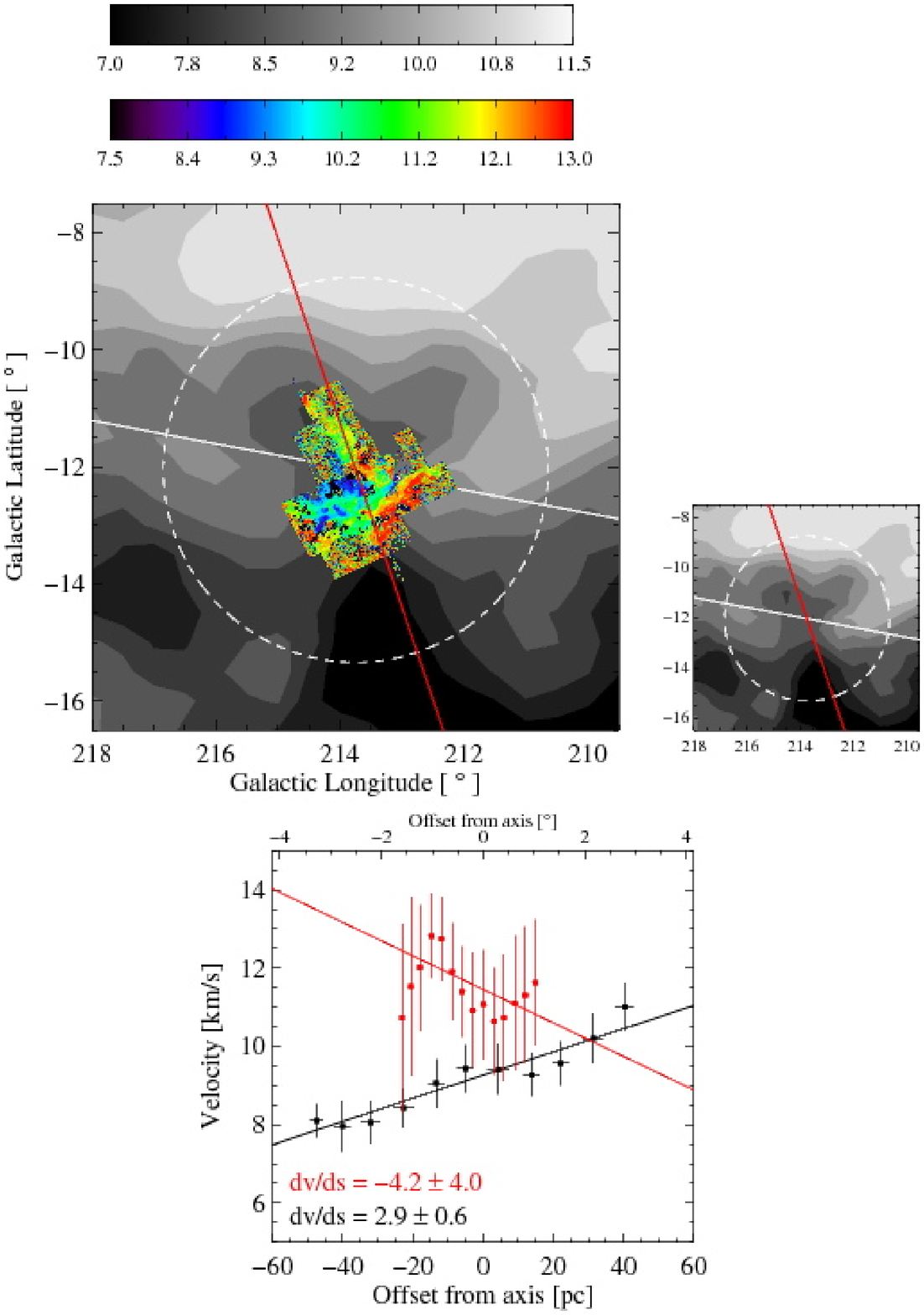}\\
\caption{MonR2.  Same as Figure \ref{fig:oriona}, except the \HI~data in the position-velocity plot are binned every 9 pc.  Although there is no significant linear gradient across the face of the GMC, we plot the ``rotation axis'' (red line) measured using Equation \ref{eq:theta} as if there was.  \label{fig:monr2}}
\end{figure*}

\begin{figure*}[htbp]
\centering
\includegraphics[scale=.65]{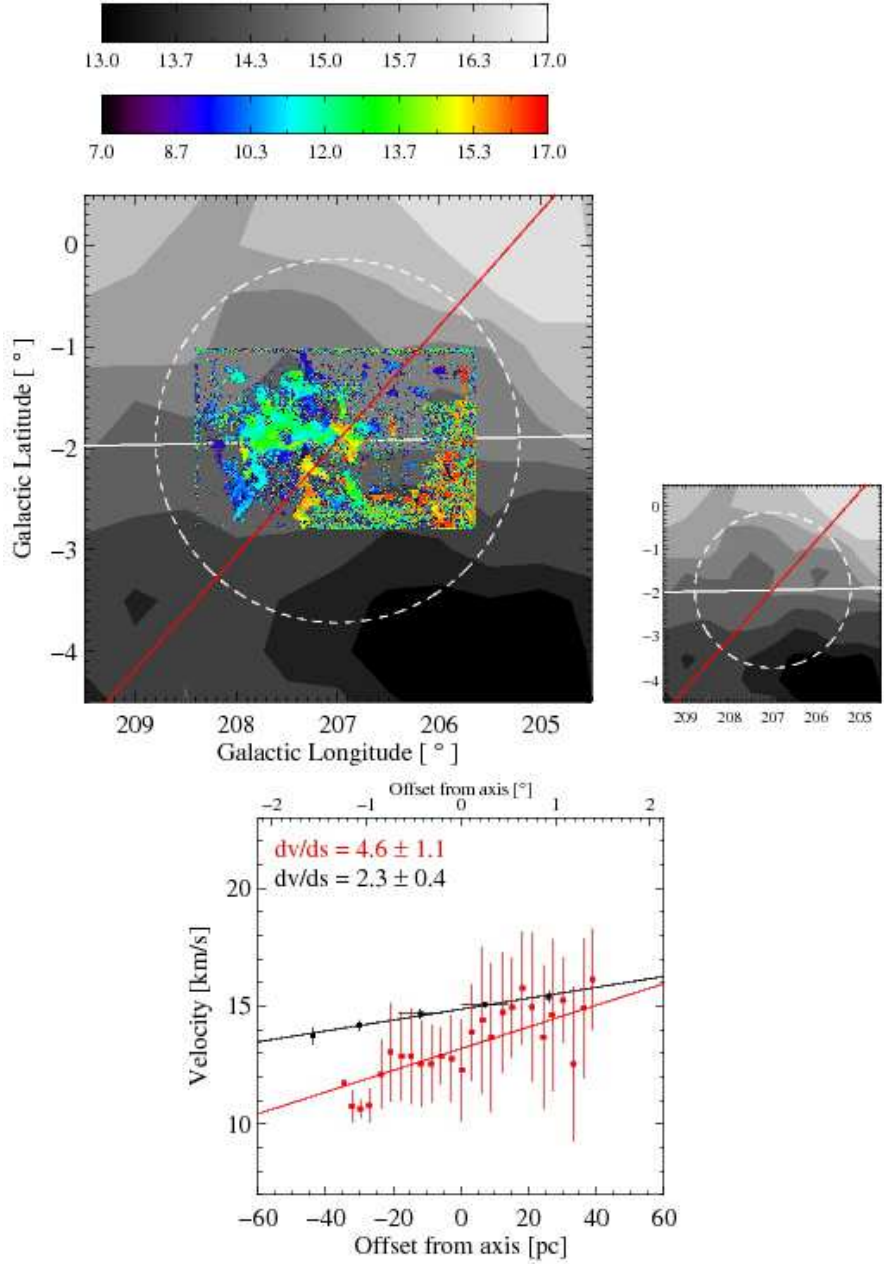}\\
\caption{Rosette.  Same as Figure \ref{fig:oriona}, except the \HI~data in the position-velocity plot are binned every 17 pc. \label{fig:rosette}}
\end{figure*}

4. We observe large differences between the velocity gradient directions in the GMCs and the atomic gas, with Orion A being the one exception.  At  more than $130^\circ$, two of the most extreme angle separations can be seen in the MonR2 and the Rosette systems  (Figure \ref{fig:monr2} and \ref{fig:rosette}).  Furthermore, the gradient directions in neither the molecular nor the atomic gas are in alignment with the overall direction of Galactic rotation.  If the gradients were due to rotation, this indicates that some GMCs are counterrotating with respect to the Galaxy.

5. That the velocity gradient position angles in the atomic and molecular gas are divergent indicates that the GMCs in our sample probably did not inherit their present velocity fields from the atomic gas from which they formed.   Finally, in at least two cases, NGC 2264 and the Rosette, a good explanation of the morphology and kinematics observed in the gas is that they are caused by stellar winds from O stars in the \HII~regions located in the GMCs, not rotation.



\begin{table*}[ht]\footnotesize\centering
\begin{tabular}{lcccccccccc}
\multicolumn{11}{c}{\textbf{TABLE 1}} \\
\multicolumn{11}{c}{\textbf{GMC Properties}} \\
\hline \hline 
          &  \multicolumn{2}{c}{\footnotesize{Coordinates}} & \footnotesize{Distance} & $M_{13}$  & $M_{\t{GMC}}$ & \footnotesize{Projected Area}  & $v_{\t{LSR}}(\co{13})$ & $v(\HI)$ & \multicolumn{2}{c}{\footnotesize{References}}  \\
 Cloud    &   $l_0$ & $b_0$ & [pc]  & [$10^4~\msun$] & [$10^4~\msun$] & [pc$^2$]& [\kms]      &   [\kms] & \footnotesize{Distance} & \co{12}    \\\hline
Perseus....... & 159 & -20  & 320   & 1.4 &  1.2 & 825   &  ~7.5 &    $-8,14$ & 1 & 6  \\
Orion A....... & 210 & -19  & 414   & 2.4 &  6.9 & 1510  & ~9.6 &    $-7,22$ &  2 & 7     \\
NGC 2264...    & 202 & 1    & 800   &  2.5 &  2.2 & 674  & ~6.6 &    $-7,15$ &  3 & 8    \\
MonR2........  & 213 & -13  &  830  & 3.0 & 9.0  & 2940  &  10.8 &    $-5,23$  &  4 & 9  \\
Rosette........& 207 & -2  &  1600  & 4.0 & 7.7 & 2200 &   13.1 &    $~4,27$  &  5 & 10   \\
\hline
\end{tabular}
\caption*{Properties are determined using the methods outlined in \S \ref{sec:analysis} of the text.  Masses calculated using \co{13} observations are listed under $M_{13}$. The GMC masses calculated by previous authors are listed under $M_\t{GMC}$.   \\
{\sc References.}--(1) de Zeeuw et al. 1999; (2) Genzel et al. 1981; (3) Sagar \& Joshi ; (4) Racine 1968; (5) Blitz \& Stark 1986; (6) Sargent 1979; (7) Menten et al. 2007; (8) Blitz 1978; (9) Maddalena et al. 1986; (10) Williams et al. 1995.}
\end{table*}

\begin{table*}[htbp!]\centering
\begin{tabular}{lcc}
\multicolumn{3}{c}{\textbf{TABLE 2}} \\
\multicolumn{3}{c}{\textbf{Accumulation Radii}} \\
\hline \hline 
Cloud          & $R_\t{GMC}$  &  $R_A$     \\
\hline
Perseus......  & 16.2   & 42.4      \\
Orion A....... & 21.9   & 76.0     \\
NGC 2264...    & 14.6   & 51.9     \\
MonR2........  & 30.6   & 83.6     \\
Rosette........& 26.5   & 78.8     \\
\hline
\end{tabular}
\caption*{The effective GMC radii, $R_\t{GMC}$, and the accumulation radii, $R_A$, are listed in units of pc. Lower limits of $R_A$ are estimated using Equation \ref{eq:accum}, assuming that the GMCs initially formed from \HI having an average surface density of 10 \sunits.   }
\end{table*}

\begin{table*}[htbp]\footnotesize\centering
\begin{tabular}{lcccccc}
\multicolumn{7}{c}{\textbf{TABLE 3}} \\
\multicolumn{7}{c}{\textbf{GMC Properties: Dynamics}} \\
\hline \hline 
Cloud &  $\Omega_\t{GMC}$ & $\Omega_\t{HI}$  & $\theta_\t{GMC}$ & $\theta_\t{HI}$  & $j_{\t{GMC}}$ & $j_\t{HI}$  \\
     &[$0.01~\vunits$] &[$0.01~\vunits$]    & [deg]          &  [deg]          & [\junits]    &   [\junits] \\\hline
Perseus.......  & 23.1     & 6.70        & -20.2     &  99.4   & 24.2   & 60.3 \\
Orion A.......  & 22.0     & 3.71        & -111      &  -102   & 42.2 [84.1]   & 107  \\
NGC 2264...     & 4.60     & 1.87        & 47.8      & -173    & ~3.92  & 25.2   \\
MonR2........   & 6.68     & 3.48        & -72.4     &  -11.2  & 25.5   & 120  \\
Rosette........ &  9.18    & 2.36        & -132      & ~1.12   & 25.8   & 73.4   \\
\hline
\end{tabular}
\caption*{The magnitude of the velocity gradient, $\Omega$, and the direction of rotation $\theta$ measured in degrees East of North are determined using the methods outlined in the text. The specific angular momentum, $j$, is calculated for each GMC and \HI region assuming that the gradients are due to rotation.  Listed in brackets is $j$ for Orion A, assuming that the GMC has the morphology of a cylinder with a rotation axis perpendicular to its long axis.    Typical errors are $\delta\Omega=0.001-0.005$, $\delta\theta(\t{CO})=0.5-2^\circ$, $\delta\theta(\t{HI})=5-10^\circ$, and $\delta j = 0.5-2~\junits$.}
\end{table*}

\clearpage

\end{document}